\documentclass[prd,showpacs,twocolumn,
amsmath,amssymb,superscriptaddress,floatfix,nofootinbib]{revtex4-1}
\usepackage[utf8]{inputenc}
\usepackage[sort&compress]{natbib}
\usepackage[normalem]{ulem}
\usepackage{lipsum} 
\usepackage{bm}
\usepackage{times}
\usepackage{amssymb,amsbsy,amsmath,amsfonts}
\usepackage{graphicx}
\usepackage{float}
\usepackage{color}
\usepackage{morefloats}
\usepackage{rotating}
\usepackage{srcltx}
\usepackage{slashed}
\usepackage{subfigure}
\usepackage{multirow}
\usepackage{verbatim}
\usepackage{hyperref}
\usepackage{tabularx}
\usepackage{adjustbox}
\usepackage[dvipsnames]{xcolor}
\usepackage{nicefrac}

\usepackage{hyperref}
\hypersetup{
	colorlinks	=true,
	urlcolor	=blue,
	linkcolor	=blue,
	citecolor	=blue,
	pdftitle	={},
	pdfauthor	={Zhi-Wei Liu, Duo-Lun Ge, Ming-Zhu Liu, Jun-Xu Lu, Li-Sheng Geng},
	pdfsubject	={Femtoscopic correlation function of the charmonium-nucleon}
	}

\begin{document}

\title{
Charmonium-nucleon femtoscopic correlation function}

\author{Zhi-Wei Liu}
\affiliation{School of Physics, Beihang University, Beijing 102206, China}

\author{Duo-Lun Ge}
\affiliation{School of Physics, Beihang University, Beijing 102206, China}

\author{Jun-Xu Lu}
\affiliation{School of Physics, Beihang University, Beijing 102206, China}

\author{Ming-Zhu Liu}
\email[Corresponding author: ]{liumz@lzu.edu.cn}
\affiliation{Frontiers Science Center for Rare Isotopes, Lanzhou University,
Lanzhou 730000, China}
\affiliation{ School of Nuclear Science and Technology, Lanzhou University, Lanzhou 730000, China}

\author{Li-Sheng Geng}
\email[Corresponding author: ]{lisheng.geng@buaa.edu.cn}
\affiliation{School of Physics, Beihang University, Beijing 102206, China}
\affiliation{
Sino-French Carbon Neutrality Research Center, \'Ecole Centrale de P\'ekin/School
of General Engineering, Beihang University, 
Beijing 100191, China}
\affiliation{Peng Huanwu Collaborative Center for Research and Education, Beihang University, Beijing 100191, China}
\affiliation{Beijing Key Laboratory of Advanced Nuclear Materials and Physics, Beihang University, Beijing 100191, China }
\affiliation{Southern Center for Nuclear-Science Theory (SCNT), Institute of Modern Physics, Chinese Academy of Sciences, Huizhou 516000, China}

\begin{abstract}

This study investigates the femtoscopic correlation functions of charmonium-nucleon pairs, utilizing the lattice QCD phase shifts provided by the HAL QCD Collaboration. A ``model-independent'' formalism is employed to transform scattering phase shifts directly into momentum correlation functions, thereby circumventing the approximations inherent in traditional methods, such as the Lednický-Lyuboshits model. The $J/\psi$-$p$ correlation functions, including spin-averaged and partial-wave results, are predicted using near-physical pion mass lattice results. The $\eta_c$-$p$ correlation function is calculated for the first time. The derived correlation functions provide critical references for future experiments, such as those at the LHC, where high-precision measurements of charmonium-nucleon correlations could unveil valuable insights into non-perturbative QCD dynamics.

\end{abstract}


\maketitle

\section{INTRODUCTION}
The strong interaction, governed by quantum chromodynamics (QCD), is responsible for binding quarks and gluons into nucleons, which constitute over 99\% of the visible mass in the universe. One of its remarkable features is asymptotic freedom, which allows perturbative calculations at high energies. In contrast, confinement at low energies poses significant challenges for first-principles studies at the hadron level. The interaction between nucleons and heavy quarkonia, such as $J/\psi$ and $\eta_c$, offers a unique window into QCD dynamics~\cite{Peskin:1979va,Bhanot:1979vb}. As an Okubo-Zweig-Iizuka (OZI)-suppressed process~\cite{Okubo:1963fa,Zweig:1964jf,Iizuka:1966fk}, this interaction is expected to occur predominantly through multiple-gluon exchanges at low energies, and, therefore, it is directly connected to the matrix element of gluon fields in the nucleon~\cite{Kharzeev:1995ij}, which contributes critically to its mass via the trace anomaly~\cite{Shifman:1978zn,Ji:1994av}. Additionally, the $J/\psi$-nucleon ($J/\psi$-$N$) interaction is necessary for understanding hidden-charm pentaquark states (e.g., the $P_c$ states observed by LHCb)~\cite{LHCb:2015yax,LHCb:2019kea} and the in-medium properties of charmonia~\cite{Sibirtsev:2005ex,Brambilla:2014jmp}.

The measurement of the $J/\psi$-$p$ invariant mass distribution~\cite{LHCb:2015yax,LHCb:2019kea} and near-threshold $J/\psi$ exclusive photoproduction off the proton~\cite{GlueX:2019mkq} have provided fundamental knowledge about the $J/\psi$-$N$ interaction. However, extracting more detailed information about the $J/\psi$-$N$ interaction remains challenging from these traditional experiments. The former lacks sufficient constraints near the threshold, while the latter's analysis remains controversial due to the on-shell assumption of $J/\psi$ and nucleon in the vector-meson dominance model~\cite{Du:2020bqj,JointPhysicsAnalysisCenter:2023qgg}. Recently, femtoscopy -- a technique that analyzes momentum correlations between particles emitted in high-energy collisions -- has emerged as a powerful alternative for probing the strong interaction~\cite{Fabbietti:2020bfg,Liu:2024uxn}. By measuring momentum correlation functions (CFs), femtoscopy has provided valuable insights into rare hadron-hadron scattering processes~\cite{STAR:2014dcy,STAR:2015kha,ALICE:2019gcn,ALICE:2019hdt,ALICE:2020mfd,ALICE:2021cpv,Si:2025eou}, which also triggered a large number of theoretical studies~\cite{Morita:2014kza,Haidenbauer:2018jvl,Kamiya:2019uiw,Liu:2022nec,Molina:2023oeu,Yan:2024aap,Achenbach:2024wgy,Li:2024tvo,Ge:2025put,Ikeno:2025kwe,Liu:2025eqw}, especially in the charm and bottom sectors~\cite{Kamiya:2022thy,Liu:2023uly,Liu:2023wfo,Vidana:2023olz,Ikeno:2023ojl,Torres-Rincon:2023qll,Feijoo:2023sfe,Khemchandani:2023xup,Li:2024tof,Albaladejo:2024lam,Liu:2024nac,Geng:2025ruq}. It is worthwhile noting that the ALICE Collaboration has recently successfully measured the $\bar{D}N$, $D^{(*)}\pi$, and $D^{(*)}K$ CFs~\cite{ALICE:2022enj,ALICE:2024bhk}, paving the way for precision studies of the strong interaction in the charm sector, such as $J/\psi$-$p$ and $\eta_c$-$p$.

The charmonium-nucleon interaction has been studied in either phenomenological models or first-principles calculations. The interaction mediated by multiple-gluon exchanges closely resembles the van der Waals force. It is, therefore, termed the gluonic van der Waals interaction~\cite{Peskin:1979va,Bhanot:1979vb,Appelquist:1978rt,Brodsky:1997gh}, which has been frequently employed to describe the charmonium-nucleon interaction. In a recent work~\cite{Wu:2024xwy}, the authors propose that $J/\psi$-$N$ scattering can occur through two distinct mechanisms: the soft-gluon exchange mechanism and the coupled-channel mechanism~\cite{Lipkin:1996ny} via open-charm meson-baryon intermediate states. According to their calculations, the soft-gluon exchange mechanism yields a scattering length~\footnote{In our convention, a negative (positive) real value indicates an attractive interaction (a repulsive interaction or the existence of a bound state).} of $<-0.16$ fm, while the coupled-channel mechanism gives a result between $-10\times10^{-3}$ and $-0.1\times10^{-3}$ fm. On the other hand, various lattice QCD collaborations have also studied the charmonium-nucleon interaction~\cite{Yokokawa:2006td,Liu:2008rza,Kawanai:2010ev,Alberti:2016dru,Sugiura:2019pye,Skerbis:2018lew}. However, due to the quenched approximation~\cite{Yokokawa:2006td,Kawanai:2010ev}, the use of heavy pion masses~\cite{Alberti:2016dru,Sugiura:2019pye}, and the significant systematic uncertainties, the early lattice simulations have not yielded consistent results. Recently, the HAL QCD Collaboration reported the first lattice simulation of the $J/\psi$-$N$ and $\eta_c$-$N$ interactions with a nearly physical pion mass ($m_\pi\approx146$ MeV)~\cite{Lyu:2024ttm}. The extracted scattering lengths, $a_{^4S_{3/2}}=-0.30(2)(^{+0}_{-2})$ fm and $a_{^2S_{1/2}}=-0.38(4)(^{+0}_{-3})$ fm, are compatible with the results derived from the soft-gluon exchange mechanism~\cite{Wu:2024xwy}, the multipole expansion and low-energy theorems in QCD~\cite{Sibirtsev:2005ex}, and the QCD sum rule analysis~\cite{Hayashigaki:1998ey}.

Bridging observables and the charmonium-nucleon interaction is a shared priority for both experimental and theoretical communities. More recently, utilizing the state-of-the-art HAL QCD potential~\cite {Wen:2025wit}, the authors studied the $J/\psi NN$ and $\eta_cNN$ systems and excluded the existence of bound states or resonances in these systems. In femtoscopy, the investigation of the $J/\psi$-$p$ CF has pioneered links between these correlations and the matter distribution in nucleons~\cite{Krein:2020yor,Krein:2022fhf,Krein:2023azg}. However, these femtoscopy studies remain limited by the use of early lattice results and the Lednický-Lyuboshits (LL) model for $J/\psi$-$p$ CF calculations. In the present work, we overcome these limitations by employing the state-of-the-art lattice QCD simulations~\cite{Lyu:2024ttm} and our recently developed ``model-independent'' formalism~\cite{Liu:2024nac}, which directly connects the CF to the scattering phase shifts (PSs). We not only re-examine the $J/\psi$-$p$ CF but also predict the $\eta_c$-$p$ CF for the first time.

This paper is organized as follows: Sect. II reviews the recently proposed formalism for the relation between CFs and  PSs. Sect. III presents predictions for $J/\psi$-$p$ and $\eta_c$-$p$ CFs by using the HAL QCD phase shifts as input and compares these results with those obtained using the LL model. Conclusions and an outlook are given in Sect. IV.

\section{THEORETICAL FRAMEWORK}
In this section, we explain how to relate PSs to CFs in a ``model-independent''~\footnote{In the present work, the ``model-independent'' refers to the strict relation between the correlation function and scattering phase shift, which is derived from Eq.~\eqref{Eq:CF_Tmatrix}. It should be emphasized that, theoretical calculations of correlation functions inevitably depend on specific treatments of the source function and the scattering wave function (e.g., $S$-wave approximation, coupled-channel effects, and off-shell effects, etc.).} way. According to the Koonin-Pratt formula~\cite{Koonin:1977fh,Pratt:1990zq}, CFs depend on two quantities: 1) the particle-emitting source created in relativistic pp, pA, and AA collisions; 2) the scattering wave function of the relative motion for the pair of interest, which contains the information on the final-state interaction and can be evaluated by using the reaction amplitude $T$-matrix.  With the above-specified ingredients, CFs are expressed as~\cite{Vidana:2023olz,Ikeno:2023ojl,Torres-Rincon:2023qll,Feijoo:2023sfe,Khemchandani:2023xup,Li:2024tof,Albaladejo:2024lam,Liu:2024nac}
\begin{align}
  C(k)&=1+\int\limits_0^\infty{\rm d}^3r~S_{12}(r)\nonumber\\
  &\times\left[\left|j_0(kr)+T(\sqrt{s})\cdot\widetilde{G}(r,\sqrt{s})\right|^2-|j_0(kr)|^2\right],\label{Eq:CF_Tmatrix}
\end{align}
where $j_0(kr)$ is the spherical Bessel function, $k=\sqrt{s-(m+M)^2}\sqrt{s-(m-M)^2}/(2\sqrt{s})$ represents the center-of-mass (c.m.) momentum of the particle pair with the charmonium mass $m$, nucleon mass $M$, and c.m. energy $\sqrt{s}$. The quantity $\widetilde{G}$ is given by
\begin{align}
  \widetilde{G}(r,\sqrt{s})=\int\limits_0^{q_{\rm max}}&\frac{{\rm d}^3k'}{(2\pi)^3}\frac{\omega(k')+E(k')}{2\omega(k')\cdot E(k')}\nonumber\\
  &\times\frac{2M\cdot j_0(k'r)}{s-[\omega(k')+E(k')]^2+i\varepsilon},\label{Eq:LFj0}
\end{align}
with $\omega(k')=\sqrt{m^2+k^{\prime2}}$ and $E(k')=\sqrt{M^2+k^{\prime2}}$. It should be noted that the derivation of Eq.~\eqref{Eq:CF_Tmatrix} starts from a separable potential $V\cdot\theta(q_{\rm max}-k)\cdot\theta(q_{\rm max}-k')$ in momentum space, which reverts into the separable scattering matrix $T\cdot\theta(q_{\rm max}-k)\cdot\theta(q_{\rm max}-k')$. When calculating the scattering wave function, the on-shell $T$-matrix and $\theta(q_{\rm max}-k)$ factor are factored out of the integrals, and $\theta(q_{\rm max}-k)$ factor is inoperative since one always works with momenta smaller than $q_{\rm amx}$. The $\theta(q_{\rm max}-k')$ factor, which controls the off-shell behavior, goes into the integral of $\widetilde{G}$ and is absorbed as the upper limit of the integral. This formalism retains the full $r$-dependence of the wave function, particularly including the short-distance contribution from the $\tilde{G}(r,s)$ function. In addition, this formalism enables the investigation of off-shell effects by varying the sharp cutoff $q_{\text{max}}$ from 0.5 to 1.5 GeV. For more details, see Refs.~\cite{Vidana:2023olz,Molina:2025lzw}. In the convention used in this work, the $T$- and $S$-matrices are related as
\begin{align}
  S(\sqrt{s})&=\exp[2i\delta(k)]\nonumber\\
  &=1-2i\rho(k)\cdot T(\sqrt{s}),\label{Eq:Smatrix}
\end{align}
where $\delta$ is the scattering PS, and $\rho(k)=2Mk/(8\pi\sqrt{s})$ is the phase-space factor. The relation between the $T$-matrix and the PS is
\begin{align}
  T(\sqrt{s})=-\frac{\exp[i\delta(k)]\cdot\sin\delta(k)}{\rho(k)}.\label{Eq:Tmatrix2}
\end{align}
From Eq.~\eqref{Eq:Tmatrix2}, one can obtain the following relation
\begin{align}
  &\left|j_0(kr)+T(\sqrt{s})\cdot\widetilde{G}(r,\sqrt{s})\right|\nonumber\\
  =&\left[j_0(kr)\cdot\cos\delta(k)-\frac{\sin\delta(k)}{\rho(k)}\cdot{\rm Re}\widetilde{G}(r,\sqrt{s})\right].\label{Eq:WF}
\end{align}
Note that ${\rm Im}\widetilde{G}(r,\sqrt{s})=-\rho(k)\cdot j_0(kr)$ is used in the above derivation. Substituting Eq.~\eqref{Eq:WF} into Eq.~\eqref{Eq:CF_Tmatrix}, one can express the CF in terms of the PS as
\begin{align}
  C(k)=1&+\mathcal{F}_1(R,k)\cdot\sin^2\delta(k)\nonumber\\
  &+\mathcal{F}_2(R,k)\cdot\sin\delta(k)\cos\delta(k),\label{Eq:CF_PS}
\end{align}
where the functions $\mathcal{F}_1(R,k)$ and $\mathcal{F}_2(R,k)$ are given by
\begin{align}
  \mathcal{F}_1(R,k)&=\int\limits_0^\infty{\rm d}^3r~S_{12}(r)\left[\left(\frac{{\rm Re}\widetilde{G}(r,\sqrt{s})}{\rho(k)}\right)^2-\left|j_0(kr)\right|^2\right],\label{Eq:F1_1}\\
  \mathcal{F}_2(R,k)&=-\int\limits_0^\infty{\rm d}^3r~S_{12}(r)\left[2j_0(kr)\cdot\frac{{\rm Re}\widetilde{G}(r,\sqrt{s})}{\rho(k)}\right].\label{Eq:F2_1}
\end{align}

It is worth emphasizing that Eqs.~\eqref{Eq:CF_PS}, \eqref{Eq:F1_1}, and \eqref{Eq:F2_1} provide a ``model-independent'' relation between CFs and PSs. It can be used not only to extract scattering parameters near a threshold but also to obtain interaction information across a broader momentum range. In contrast to the widely used LL model, this formalism eliminates approximations that replace realistic wave functions with asymptotic wave functions and corrections based on effective ranges. Such improvements prevent large deviations in scenarios involving smaller source sizes, as discussed in the next section and Ref.~\cite{Liu:2024nac}. It is well known that different forms of nuclear potentials, such as the Argonne V$_{18}$ potential~\cite{Wiringa:1994wb}, the CD-Bonn potential~\cite{Machleidt:2000ge}, and chiral potentials~\cite{Epelbaum:2014sza,Lu:2021gsb}, can reproduce identical nucleon-nucleon PSs~\cite{Lu:2025syk}. However, subtle differences in these realistic nuclear forces can lead to significant divergences in nuclear many-body calculations~\cite{Epelbaum:2008ga,Machleidt:2011zz,Hammer:2019poc}. This suggests that converting CFs into PSs (or vice versa) may hold more general significance than directly constraining unknown potentials through CFs.

In the derivation of Eqs.~\eqref{Eq:F1_1} and \eqref{Eq:F2_1}, we did not specify the explicit form of the source function. Numerous methods have been developed to eliminate the ambiguity induced by source functions. A prominent example is the resonance source model, a data-driven approach proposed by the ALICE Collaboration, which has been widely adopted in femtoscopic studies~\cite{ALICE:2020mfd, ALICE:2021cpv, ALICE:2022enj}. In this model, the source is characterized by a Gaussian core that emits all primordial particles and exhibits a clear transverse mass ($m_{\rm T}$) scaling~\cite{ALICE:2020ibs, ALICE:2023sjd}, complemented by an exponential tail arising from strongly decaying resonances. Meanwhile, they found that this two-component source can be effectively replaced by a single Gaussian source~\cite{ALICE:2020ibs}. Albeit losing the direct physical interpretation of the source size, the effective Gaussian source has been successfully used in experimental analyses~\cite{ALICE:2019hdt, ALICE:2020mfd}. Furthermore, our previous work~\cite{Liu:2022nec} systematically compared Gaussian and Cauchy (exponential-tailed) source profiles using $\Xi^-p$ correlation data. The results showed negligible sensitivity to the source shape within current experimental uncertainties. Recent advancements include the optical deblurring algorithm for imaging the source in heavy-ion collisions~\cite{Xu:2024dnd} and the reconstruction of proton-emitting sources from experimental CFs using deep neural networks in an automatic differentiation framework~\cite{Xu:2024dnd,Wang:2024bpl}. The above sophisticated source functions can be easily integrated into Eqs.~\eqref{Eq:F1_1} and \eqref{Eq:F2_1}. To facilitate the practical application of Eq.~\eqref{Eq:CF_PS} and the comparison with the LL model, we comment that adopting a Gaussian source with a single parameter $R$, namely, $S_{12}(r)=\exp\left[-r^2/(4R^2)\right]/(2\sqrt{\pi}R)^3$, can be a very good starting point in the present exploratory study.

\section{RESULTS AND DISCUSSIONS}
We begin by briefly reviewing the fundamental inputs of the present work -- PSs derived from the latest lattice QCD simulations by the HAL QCD Collaboration~\cite{Lyu:2024ttm}. Specifically, the simulations provide the $J/\psi$-$N$ PSs for the $^2S_{1/2}$ and $^4S_{3/2}$ partial waves, as well as the $\eta_c$-$N$ PSs for the $^2S_{1/2}$ partial wave. These PSs gradually increase with the c.m. kinetic energy ($E_{\rm c.m.}=\sqrt{m^2+k^2}+\sqrt{M^2+k^2}-m-M$) and stabilize at approximately $14^\circ$, $11^\circ$, and $9^\circ$ for the $J/\psi$-$N$ $^2S_{1/2}$, $^4S_{3/2}$, and $\eta_c$-$N$ $^2S_{1/2}$ partial waves, respectively, beyond $E_{\rm c.m.}\approx$ 30 MeV. Given that the simulations were performed for a nearly physical pion mass ($m_\pi\approx$ 146 MeV), we assume that the difference between these results and the physical ones can be neglected. In addition, due to the heavy-quark spin symmetry, the coupled-channel effect between the $J/\psi$-$N$ and $\eta_c$-$N$ systems in the $^2S_{1/2}$ partial wave is significantly suppressed. Meanwhile, for low-momentum scattering processes near the threshold, the influence of far-away channels, such as $\Lambda_c\bar{D}^{(*)}$ and $\Sigma_c\bar{D}^{(*)}$, can be safely neglected. Therefore, the HAL QCD Collaboration conducted their lattice simulations in a single-channel framework. Following the same considerations, we also adopt the single-channel approximation in our CF calculations. In addition, higher partial waves in the charmonia-proton system also contribute to the correlation function, but these contributions experience significant suppression in both the source function and final-state interactions compared to the dominant $S$-wave contribution.

\begin{figure}[htbp]
  \centering
  \includegraphics[width=0.38\textwidth]{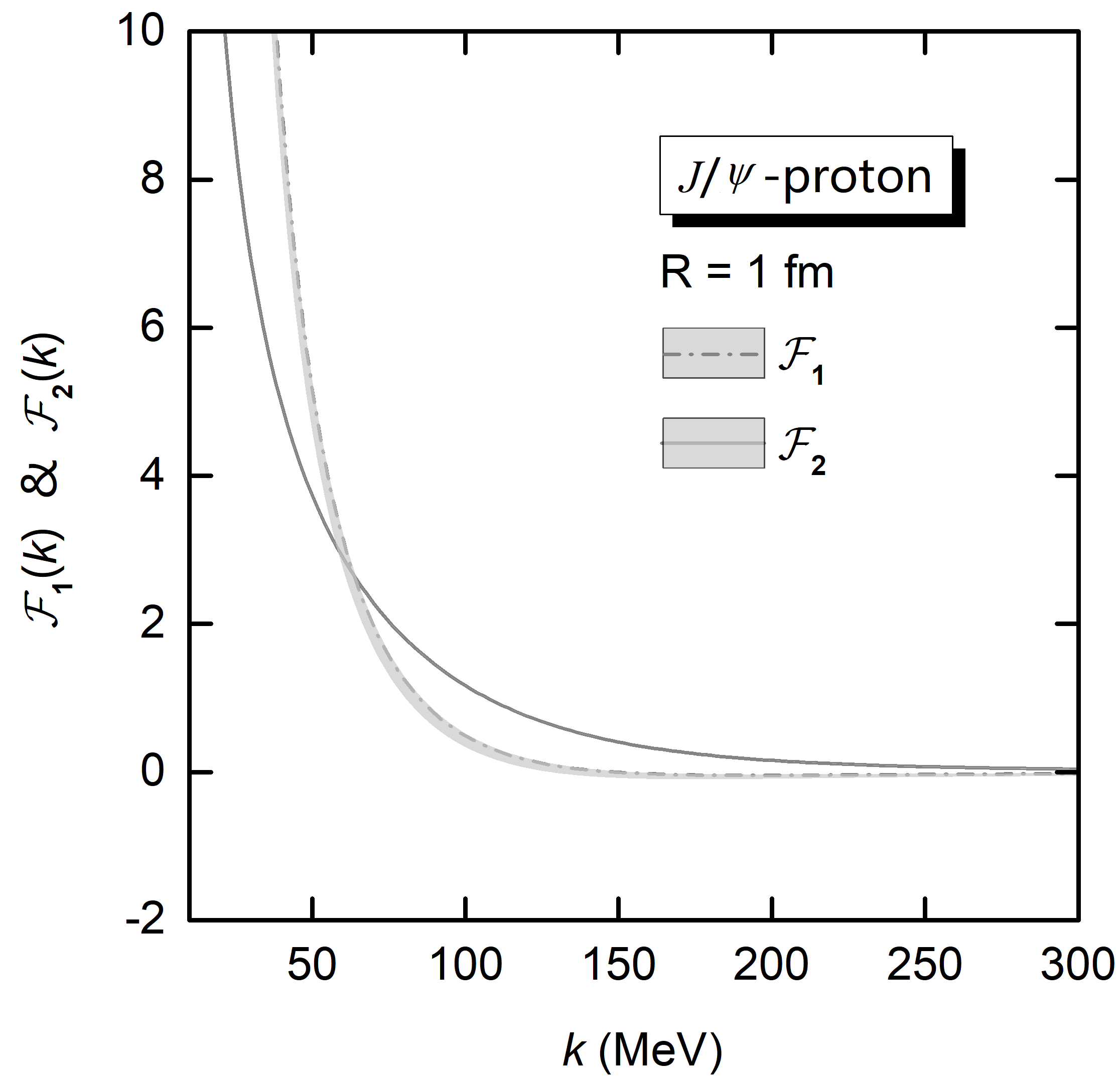}
  \caption{Functions $\mathcal{F}_1$ and $\mathcal{F}_2$ for $R = 1$ fm as a function of the relative momentum $k$. The bands reflect the variation of the sharp cutoff in the range of $q_{\rm max} = 0.5-1.5$ GeV.}\label{Fig:F1F2}
\end{figure}

Before applying Eq.~\eqref{Eq:CF_PS} to calculate CFs, it is necessary to discuss the fundamental properties of functions $\mathcal{F}_1$ and $\mathcal{F}_2$. These two functions incorporate two-body kinematic characteristics and emission source information, remaining independent of interaction dynamics. Here, we take the $J/\psi$-$p$ system and a source size of $R = 1$ fm as an example. As shown in Fig.~\ref{Fig:F1F2}, both $\mathcal{F}_1$ and $\mathcal{F}_2$ decrease rapidly with increasing relative momentum $k$, asymptotically approaching zero, which ensures the convergence of CFs to unity in the high-momentum region. It is worthwhile to note that $\mathcal{F}_2$ remains positive throughout the momentum range, implying that when the PS exceeds $90^\circ$ the term $\mathcal{F}_2\sin\delta\cos\delta$ predominantly contributes to the negative correlations -- a behavior observed in bound-state or near-threshold narrow-resonance scenarios~\cite{Liu:2024nac}. Moreover, both $\mathcal{F}_1$ and $\mathcal{F}_2$ are insensitive to the cutoff $q_{\rm max}$ value employed due to the modulating role of the spherical Bessel function $j_0(kr)$~\cite{Feijoo:2024bvn}. Compared to the uncertainties from the HAL QCD PSs (statistical errors), the cutoff variation introduces negligible effects and is thus ignored in subsequent calculations.

\begin{figure}[htbp]
  \centering
  \includegraphics[width=0.38\textwidth]{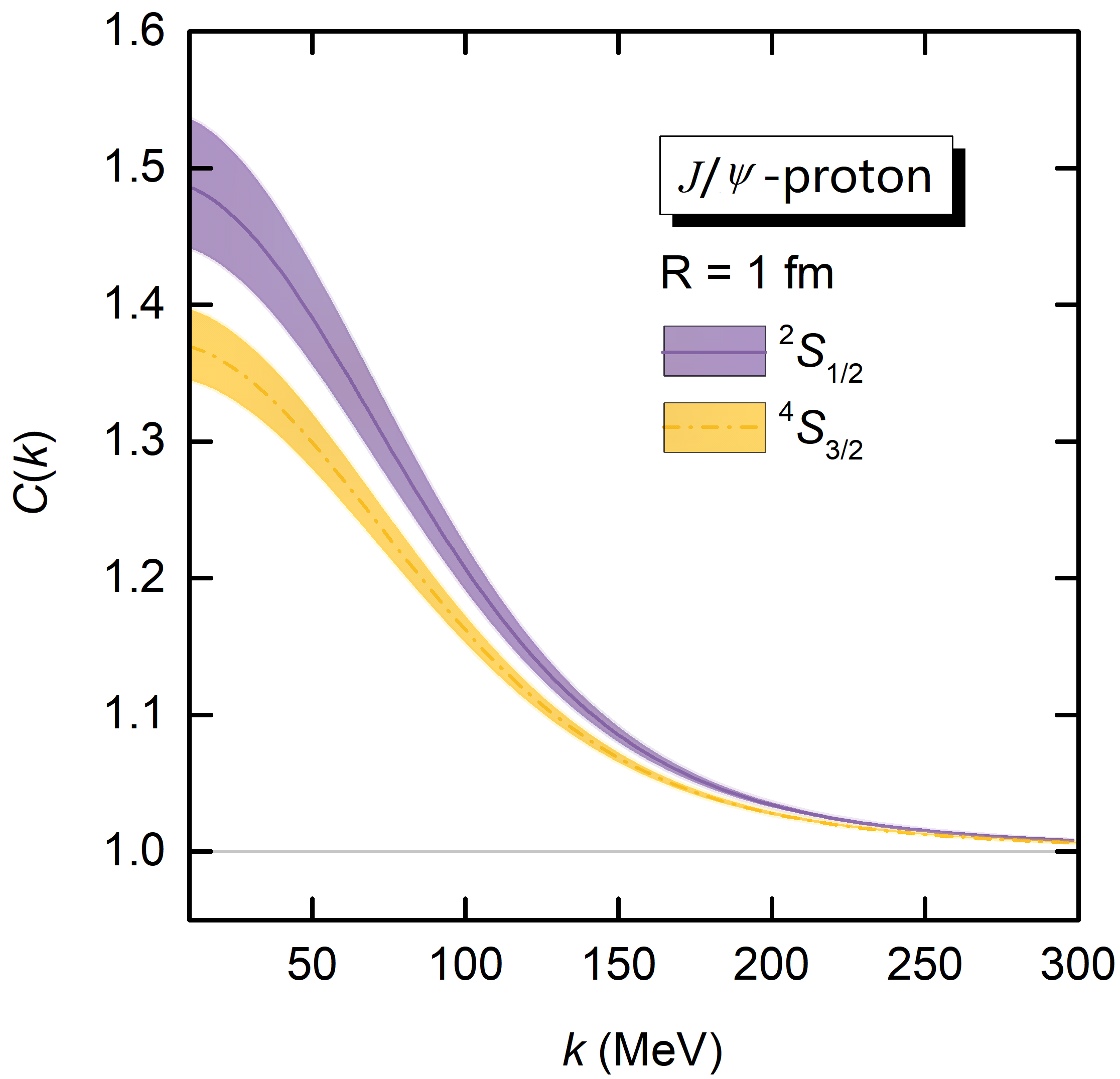}
  \caption{Spin doublet and quartet components of the $J/\psi$-$p$ correlation function as a function of the relative momentum $k$ for a source size $R = 1$ fm.}\label{Fig:spin}
\end{figure}

Next, we re-examine the $J/\psi$-$p$ CFs based on Eq.~\eqref{Eq:CF_PS} and the HAL QCD $J/\psi$-$N$ PSs. As shown in Fig.~\ref{Fig:spin}, the $J/\psi$-$p$ CFs for the $^2S_{1/2}$ and $^4S_{3/2}$ partial waves are presented for a source size $R = 1$ fm, where the shaded bands show the variation due to the statistical errors of the HAL QCD PSs. Both partial-wave results exhibit positive correlations (enhancement above unity) over a wide range of the relative momentum $k$, with the $^2S_{1/2}$ partial wave CF slightly exceeding that of the $^4S_{3/2}$ partial wave. This behavior reflects the fundamental features of two-body CFs, where attractive interactions (not strong enough to generate a bound state or a resonance) produce positive correlations across the entire momentum region, and the correlation strength increases with the intensity of the attractive interaction. 

Given the similarities between the $J/\psi$-$N$ and $\phi$-$N$ interactions, it is essential to compare the $J/\psi$-$p$ and $\phi$-$p$ CFs for two partial waves. For the $^4S_{3/2}$ partial wave, the $\phi$-$p$ CF behaves similar to the $J/\psi$-$p$ CF but with a stronger correlation strength (see Fig. 2 of Ref.~\cite{Chizzali:2022pjd}), consistent with the HAL QCD simulations showing larger $\phi$-$N$ PSs (reaching approximately $30^\circ$ at $E_{\rm c.m.}\approx$ 30 MeV~\cite{Lyu:2022imf}). While the HAL QCD Collaboration has not studied the $\phi$-$N$ interaction in the $^2S_{1/2}$ partial wave, based on the fixed $^4S_{3/2}$ interaction, the authors of Ref.~\cite{Chizzali:2022pjd} extracted the $^2S_{1/2}$ CF from the spin-averaged $\phi$-$p$ correlation data measured at $\sqrt{s} = 13$ TeV pp collisions~\cite{ALICE:2021cpv}. Unlike the $J/\psi$-$p$ $^2S_{1/2}$ CF, the obtained $\phi$-$p$ $^2S_{1/2}$ CF shows a negative correlation (reduction below unity) over a wide range of $k$, implying the appearance of a $\phi$-$p$ bound state in this partial wave.

\begin{figure}[htbp]
  \centering
  \includegraphics[width=0.38\textwidth]{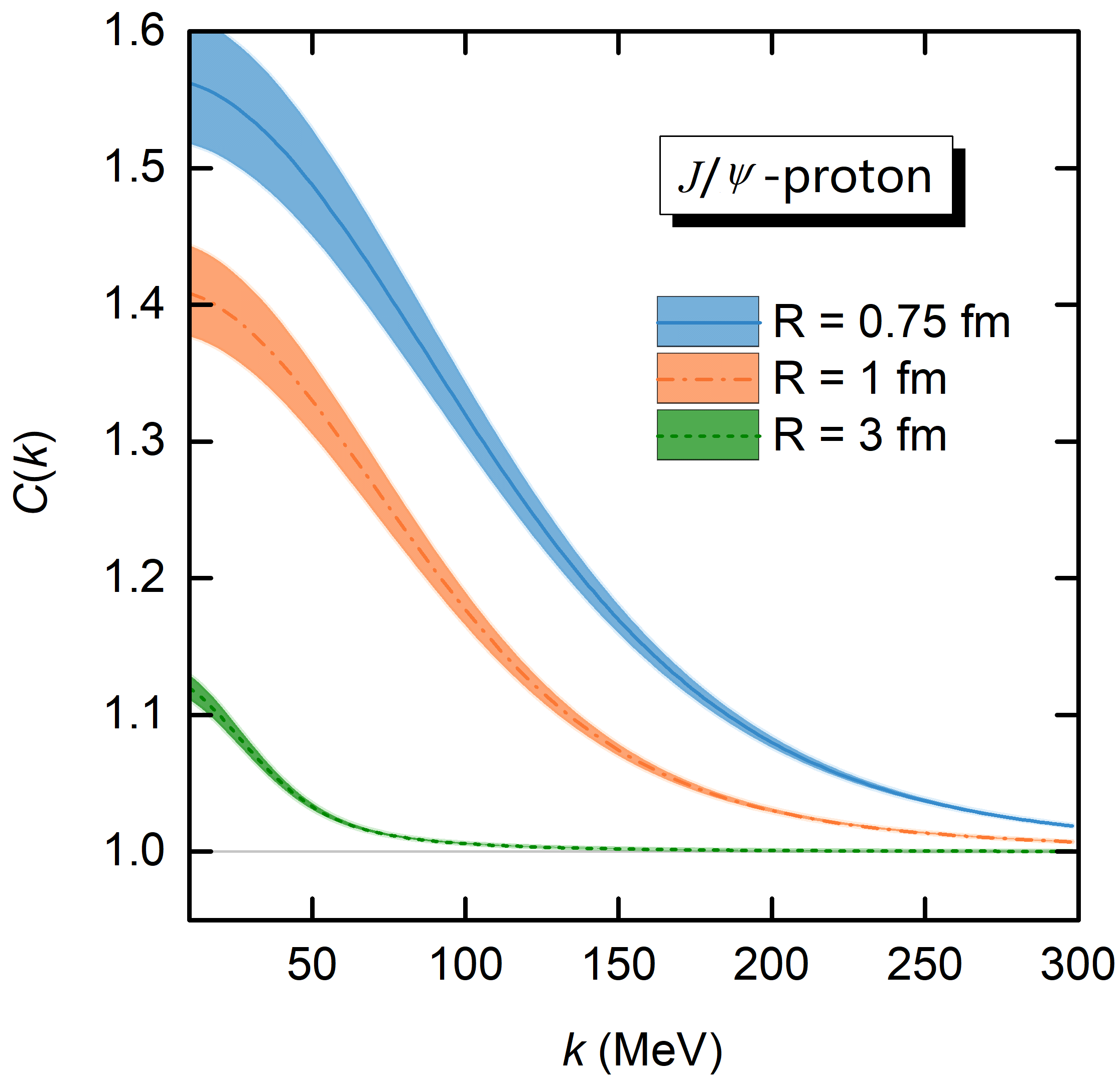}
  \includegraphics[width=0.38\textwidth]{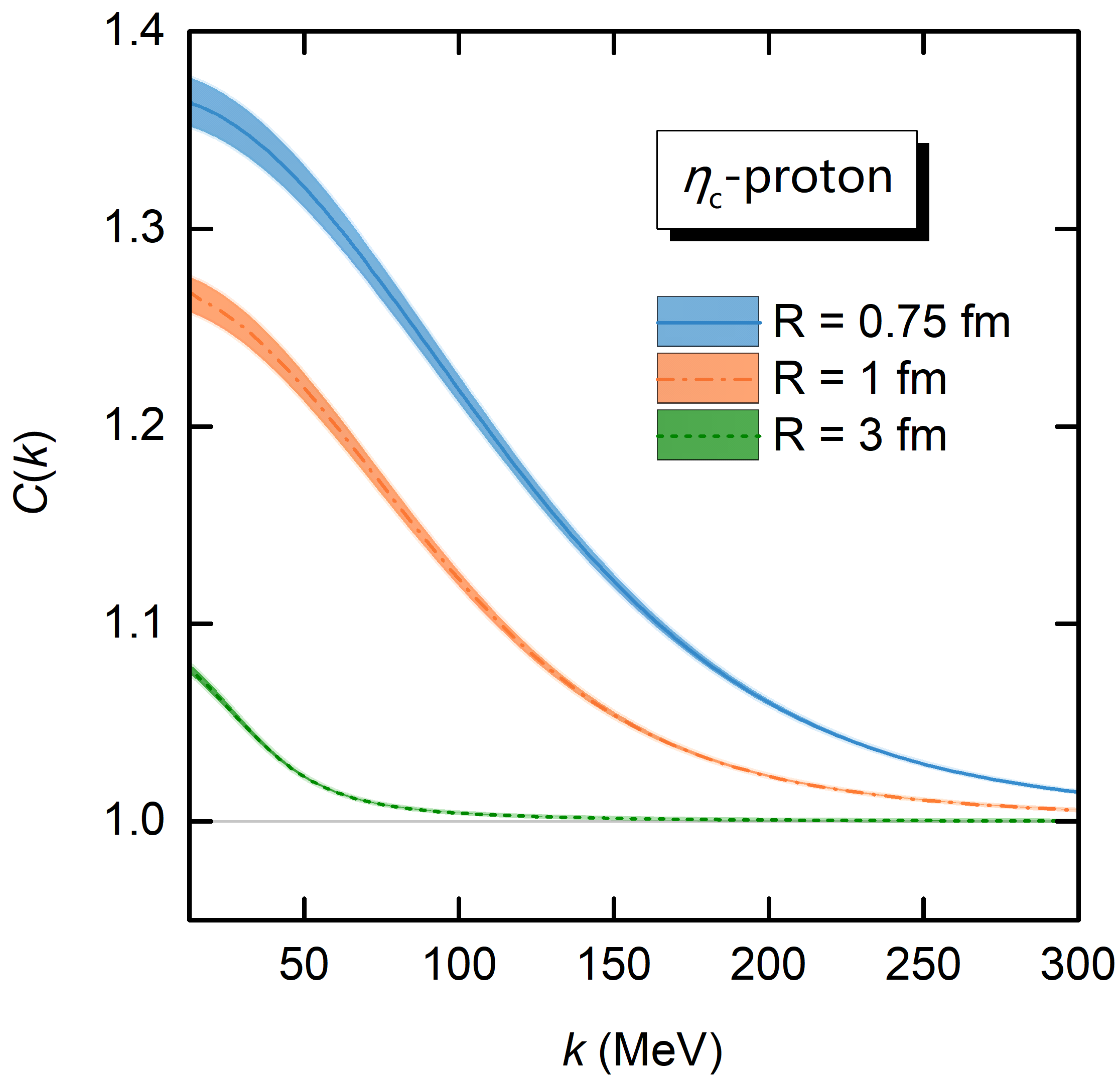}
  \caption{Spin-averaged $J/\psi$-$p$ (upper) and $\eta_c$-$p$ (lower) correlation functions as a function of the relative momentum $k$ for different source sizes $R = 0.75$, $1$, and $3$ fm.}\label{Fig:JPsip_etap}
\end{figure}

Since the spin configuration of the $J/\psi$-$p$ pair cannot be distinguished in current measurements of CFs, we have to take the spin average $C_{\rm avg}(k)=1/3~C_{^2S_{1/2}}(k)+2/3~C_{^4S_{3/2}}(k)$ to compare with future experiments. Fig.~\ref{Fig:JPsip_etap} displays these spin-averaged $J/\psi$-$p$ CFs for different source sizes $R = 0.75, 1$, and $3$ fm, where the shaded bands indicate the uncertainties from the HAL QCD PSs. Compared with the previous $R = 1$ fm results (see Fig. 2 of Ref.~\cite{Krein:2020yor}), which are calculated using the effective potential derived from the energy density and pressure inside the proton~\cite{Eides:2017xnt}, our results support the scenario of the $J/\psi$ chromopolarizability with $\alpha_{J/\psi} = 1.6$ GeV$^{-3}$. In addition, we present the predicted $\eta_c$-$p$ CFs for the first time. Given that the lattice simulations indicate a weaker attractive $\eta_c$-$N$ interaction than the $J/\psi$-$N$ interaction, the predicted $\eta_c$-$p$ CFs are slightly lower than their $J/\psi$-$p$ counterparts. These spin-averaged $J/\psi$-$p$ and $\eta_c$-$p$ CFs can be directly compared with future experimental measurements from pp, pA, and AA collisions.

In the following, we compare the $\eta_c$-$p$ CFs calculated using Eq.~\eqref{Eq:CF_PS} with the results from the LL model~\cite{Lednicky:1981su}, which has been widely used in the experimental and theoretical studies of CFs~\cite{ExHIC:2017smd, Fabbietti:2020bfg}. For systems with two non-identical particles and without the Coulomb interaction, the CF in the so-called LL model can be expressed as
\begin{align}\label{Eq:LL}
  C_{\rm LL}(k)=1&+\frac{\left|f(k)\right|^2}{2R^2}+\frac{2{\rm Re}f(k)}{\sqrt{\pi}R}F_1(x)-\frac{{\rm Im}f(k)}{R}F_2(x)\nonumber\\
  &-\frac{\left|f(k)\right|^2}{4\sqrt{\pi}R^3}r_{\rm eff},
\end{align}
where the functions $F_1(x)=\int_0^x~{\rm d}t~\exp(t^2-x^2)/x$, $F_2(x)=[1-\exp(-x^2)]/x$, and $x=2kR$. Here, the scattering amplitude $f(k)$ is given by the effective range expansion
\begin{align}\label{Eq:ERE}
  f(k)\approx(-1/a_0+r_{\rm eff}k^2/2-ik)^{-1},
\end{align}
where $a_0$ and $r_{\rm eff}$ are the scattering length and effective range, respectively. It is worthwhile emphasizing that the last term in Eq.~\eqref{Eq:LL} (the second line), the effective range correction term, is introduced to address systematic errors induced by approximating a full wave function with its asymptotic form. The derivation of this term requires that $r_{\rm eff}$ be much smaller than $R$, a condition that appears to have received insufficient attention in practice. For instance, the studies of measured $\phi$-$p$ CFs with the LL model yield a spin-averaged $r_{\rm eff} = 7.85\pm1.54({\rm stat})\pm0.26({\rm syst})$ fm, which appears in tension with the theoretical expectation $r_{\rm eff}\ll R$ given the extracted $R = 1.08\pm0.05$ fm~\cite{ALICE:2021cpv}. 

\begin{figure}[htbp]
  \centering
  \includegraphics[width=0.40\textwidth]{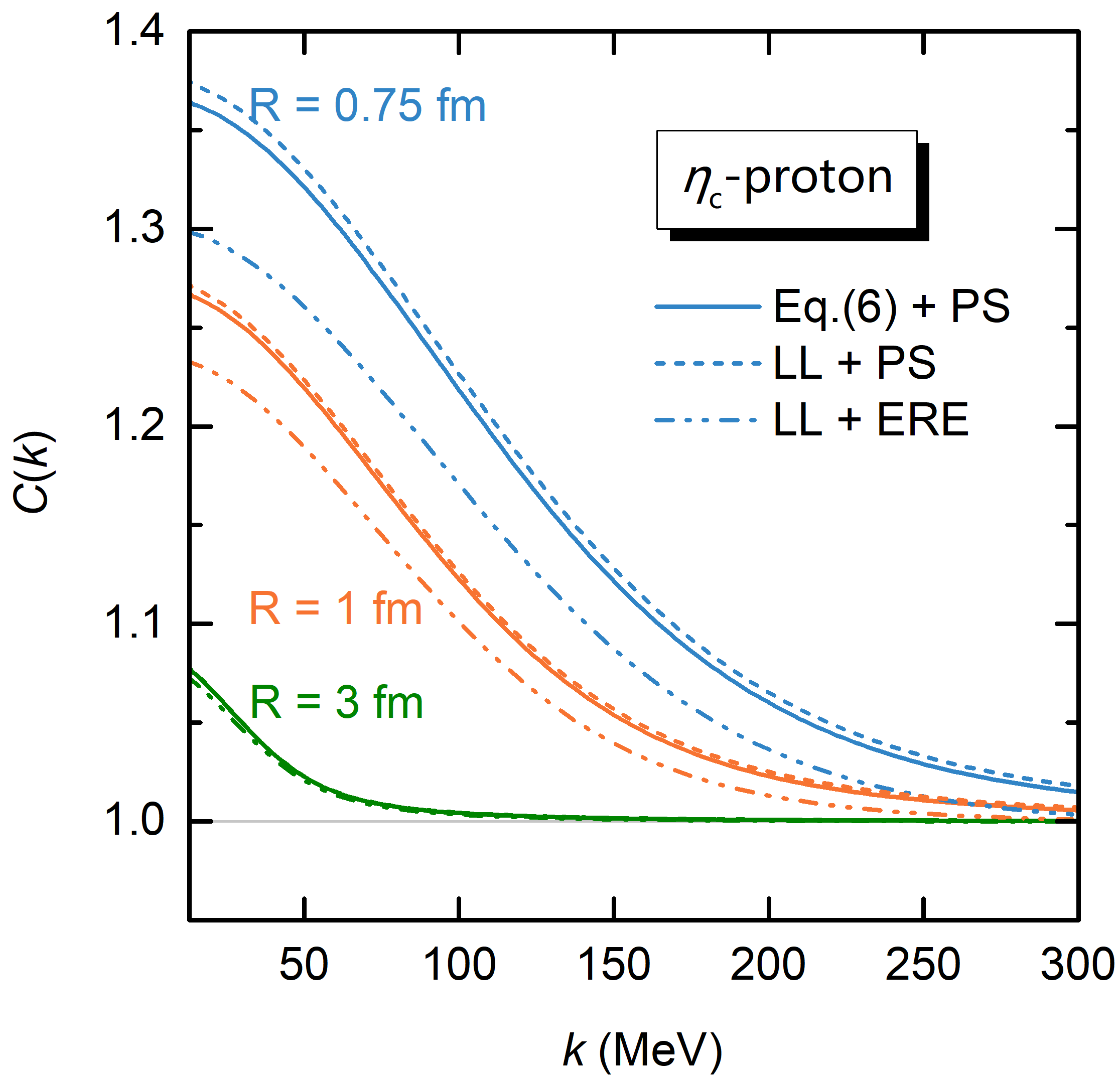}
  \caption{The $\eta_c$-proton correlation function computed using Eq.~(6) in comparison with the Lednick${\rm \acute{y}}$-Lyuboshits model result. The dash-dot-dotted lines denote the complete LL model results with the scattering parameters as input. The short-dashed lines denote the LL model results without the effective range correction term and with the phase shifts as input.}\label{Fig:LL}
\end{figure}

As shown in Fig.~\ref{Fig:LL}, we present the $\eta_c$-$p$ CFs computed using Eq.~\eqref{Eq:CF_PS} alongside those derived from the LL model. Here, we adopt the HAL QCD Collaboration's $\eta_c$-$N$ scattering parameters $a_0 = -0.21$ fm and $r_{\rm eff} = 3.65$ fm for the LL model calculations. It is seen that the LL model results are systematically lower than those from Eq.~\eqref{Eq:CF_PS}, which can be easily deduced from the negative sign of the effective range correction term. This deviation becomes more pronounced as the source size decreases. In the low-momentum region, this deviation can reach $0.07$ for $R = 0.75$ fm. When the effective range correction term is removed, the LL model results are close to but slightly higher than those in our formalism. Since both methods use the same phase shifts as input, the observed difference between the solid and dashed lines represents a systematic difference between the two methods, arising essentially from their different treatments of the wave function's short-range behavior. Therefore, we suggest employing the rigorous relation between PSs and CFs to analyze charmonium-nucleon CFs in the future.

\begin{figure}[htbp]
  \centering
  \includegraphics[width=0.40\textwidth]{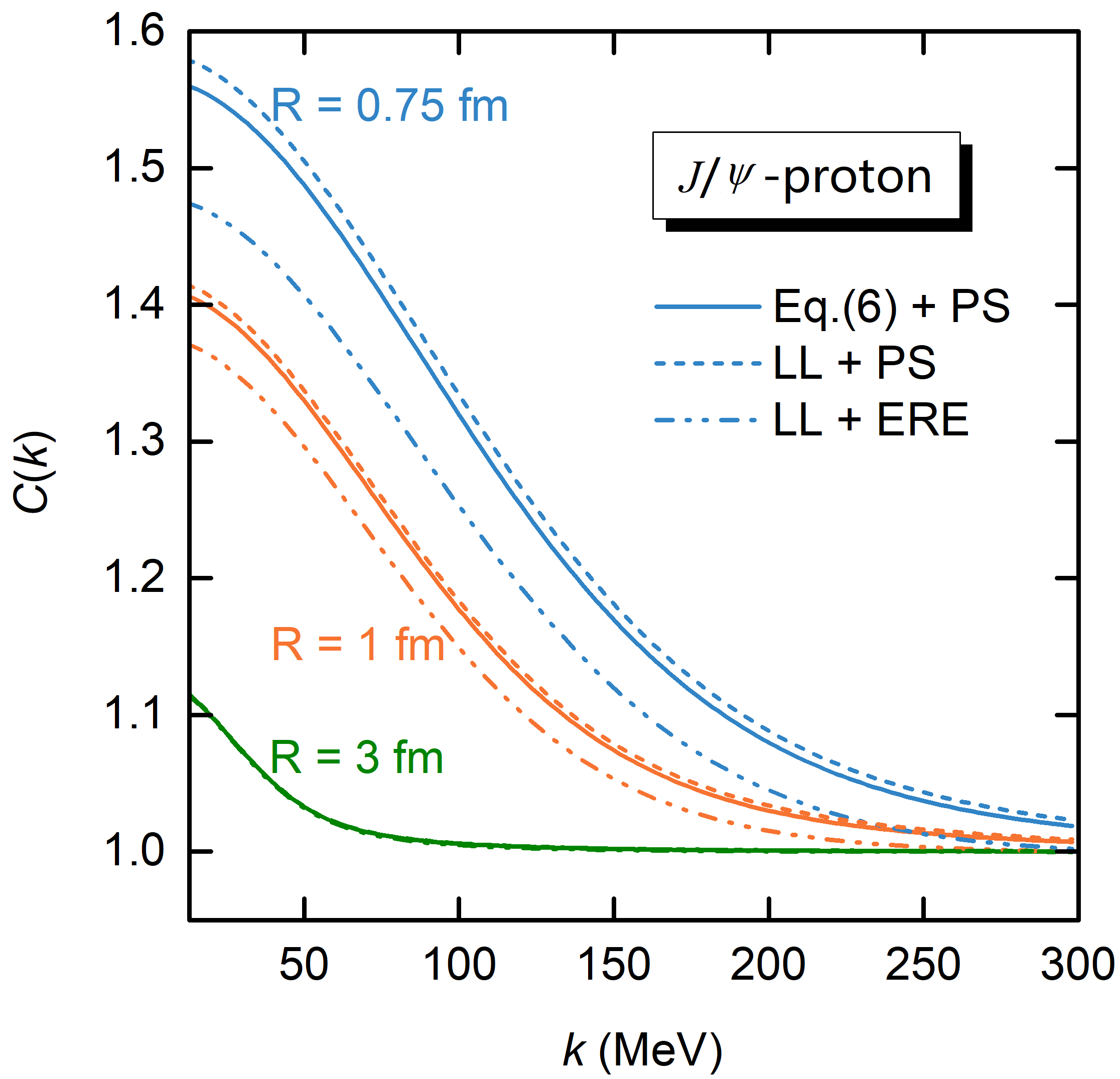}
  \caption{The spin-averaged $J/\psi$-proton correlation function computed using Eq.~(6) in comparison with the Lednick${\rm \acute{y}}$-Lyuboshits model result. The dash-dot-dotted lines denote the complete LL model results with the scattering parameters as input. The short-dashed lines denote the LL model results without the effective range correction term and with the phase shifts as input.}\label{Fig:CF_Jpsip}
\end{figure}

In Fig.~\ref{Fig:CF_Jpsip}, we present the $J/\psi$-proton correlation functions obtained from our formalism, the LL model (without the effective range correction term) using the phase shifts as input, and the LL model using the scattering parameters as input. In all three cases, the lineshapes exhibit behavior similar to those of the $\eta_c$-proton correlation functions, reflecting the similar magnitude of the $J/\psi-N$ and $\eta_c$-$N$ interactions.

\section{SUMMARY AND OUTLOOK}
In this work, we investigated the charmonium-nucleon femtoscopic correlation functions, which can be used to probe the non-perturbative interactions between charmonia and nucleons. First, we reviewed a ``model-independent'' relation between two-body scattering phase shifts and momentum correlation functions. Using the latest HAL QCD $J/\psi$-$N$ phase shifts for the $^2S_{1/2}$ and $^4S_{3/2}$ partial waves as input, we studied the $J/\psi$-$p$ correlation functions for these two partial waves, as well as the spin-averaged results, for different source sizes. Furthermore, using the $\eta_c$-$N$ HAL QCD phase shifts, we predicted the $\eta_c$-$p$ correlation function for the first time. These predictions provided direct references for future femtoscopy experiments. Additionally, comparing the Lednick${\rm \acute{y}}$-Lyuboshits model calculations with our results, we found that the effective range correction term in the Lednick${\rm \acute{y}}$-Lyuboshits model introduces non-negligible contributions as the effective range exceeds the source size. Therefore, we recommend adopting the ``phase shift to correlation function" relation for analyzing future experimental data.

The $J/\psi$ particle can be produced abundantly at the LHC~\cite{LHCb:2020bwg,ALICE:2023hou} and measured in the electromagnetic decay channels $e^+e^-$ and $\mu^+\mu^-$, each having branching ratios of approximately 6\%~\cite{ParticleDataGroup:2024cfk}. These advantages of high production rates and excellent detectability are ideal for measuring the $J/\psi$-$p$ CF. With the upgraded ALICE apparatus and the larger data sample expected at LHC runs 3, 4, and 5~\cite{ALICE:2022wwr}, we anticipate that the $J/\psi$-$p$ and $\eta_c$-$p$ CFs can be measured in the near future, providing valuable information about the charmonium-nucleon interaction.

\emph{Acknowledgments.} This work is partly supported by the National Key R\&D Program of China under Grant No. 2023YFA1606703 and the National Natural Science Foundation of China under Grant No. 12435007. ZWL acknowledges support from the National Natural Science Foundation of China under Grant Nos. 12405133 and 12347180, the China Postdoctoral Science Foundation under Grant No. 2023M740189, and the Postdoctoral Fellowship Program of CPSF under Grant No.GZC20233381.

\bibliography{JPsiN.bib}

\begin{thebibliography}{90}%
\makeatletter
\providecommand \@ifxundefined [1]{%
 \@ifx{#1\undefined}
}%
\providecommand \@ifnum [1]{%
 \ifnum #1\expandafter \@firstoftwo
 \else \expandafter \@secondoftwo
 \fi
}%
\providecommand \@ifx [1]{%
 \ifx #1\expandafter \@firstoftwo
 \else \expandafter \@secondoftwo
 \fi
}%
\providecommand \natexlab [1]{#1}%
\providecommand \enquote  [1]{``#1''}%
\providecommand \bibnamefont  [1]{#1}%
\providecommand \bibfnamefont [1]{#1}%
\providecommand \citenamefont [1]{#1}%
\providecommand \href@noop [0]{\@secondoftwo}%
\providecommand \href [0]{\begingroup \@sanitize@url \@href}%
\providecommand \@href[1]{\@@startlink{#1}\@@href}%
\providecommand \@@href[1]{\endgroup#1\@@endlink}%
\providecommand \@sanitize@url [0]{\catcode `\\12\catcode `\$12\catcode `\&12\catcode `\#12\catcode `\^12\catcode `\_12\catcode `\%12\relax}%
\providecommand \@@startlink[1]{}%
\providecommand \@@endlink[0]{}%
\providecommand \url  [0]{\begingroup\@sanitize@url \@url }%
\providecommand \@url [1]{\endgroup\@href {#1}{\urlprefix }}%
\providecommand \urlprefix  [0]{URL }%
\providecommand \Eprint [0]{\href }%
\providecommand \doibase [0]{http://dx.doi.org/}%
\providecommand \selectlanguage [0]{\@gobble}%
\providecommand \bibinfo  [0]{\@secondoftwo}%
\providecommand \bibfield  [0]{\@secondoftwo}%
\providecommand \translation [1]{[#1]}%
\providecommand \BibitemOpen [0]{}%
\providecommand \bibitemStop [0]{}%
\providecommand \bibitemNoStop [0]{.\EOS\space}%
\providecommand \EOS [0]{\spacefactor3000\relax}%
\providecommand \BibitemShut  [1]{\csname bibitem#1\endcsname}%
\let\auto@bib@innerbib\@empty
\bibitem [{\citenamefont {Peskin}(1979)}]{Peskin:1979va}%
  \BibitemOpen
  \bibfield  {author} {\bibinfo {author} {\bibfnamefont {M.~E.}\ \bibnamefont {Peskin}},\ }\href {\doibase 10.1016/0550-3213(79)90199-8} {\bibfield  {journal} {\bibinfo  {journal} {Nucl. Phys. B}\ }\textbf {\bibinfo {volume} {156}},\ \bibinfo {pages} {365} (\bibinfo {year} {1979})}\BibitemShut {NoStop}%
\bibitem [{\citenamefont {Bhanot}\ and\ \citenamefont {Peskin}(1979)}]{Bhanot:1979vb}%
  \BibitemOpen
  \bibfield  {author} {\bibinfo {author} {\bibfnamefont {G.}~\bibnamefont {Bhanot}}\ and\ \bibinfo {author} {\bibfnamefont {M.~E.}\ \bibnamefont {Peskin}},\ }\href {\doibase 10.1016/0550-3213(79)90200-1} {\bibfield  {journal} {\bibinfo  {journal} {Nucl. Phys. B}\ }\textbf {\bibinfo {volume} {156}},\ \bibinfo {pages} {391} (\bibinfo {year} {1979})}\BibitemShut {NoStop}%
\bibitem [{\citenamefont {Okubo}(1963)}]{Okubo:1963fa}%
  \BibitemOpen
  \bibfield  {author} {\bibinfo {author} {\bibfnamefont {S.}~\bibnamefont {Okubo}},\ }\href {\doibase 10.1016/S0375-9601(63)92548-9} {\bibfield  {journal} {\bibinfo  {journal} {Phys. Lett.}\ }\textbf {\bibinfo {volume} {5}},\ \bibinfo {pages} {165} (\bibinfo {year} {1963})}\BibitemShut {NoStop}%
\bibitem [{\citenamefont {Zweig}(1964)}]{Zweig:1964jf}%
  \BibitemOpen
  \bibfield  {author} {\bibinfo {author} {\bibfnamefont {G.}~\bibnamefont {Zweig}},\ }\enquote {\bibinfo {title} {{An SU(3) model for strong interaction symmetry and its breaking. Version 2}},}\ in\ \href {\doibase 10.17181/CERN-TH-412} {\emph {\bibinfo {booktitle} {{DEVELOPMENTS IN THE QUARK THEORY OF HADRONS. VOL. 1. 1964 - 1978}}}},\ \bibinfo {editor} {edited by\ \bibinfo {editor} {\bibfnamefont {D.~B.}\ \bibnamefont {Lichtenberg}}\ and\ \bibinfo {editor} {\bibfnamefont {S.~P.}\ \bibnamefont {Rosen}}}\ (\bibinfo {year} {1964})\ pp.\ \bibinfo {pages} {22--101}\BibitemShut {NoStop}%
\bibitem [{\citenamefont {Iizuka}(1966)}]{Iizuka:1966fk}%
  \BibitemOpen
  \bibfield  {author} {\bibinfo {author} {\bibfnamefont {J.}~\bibnamefont {Iizuka}},\ }\href {\doibase 10.1143/PTPS.37.21} {\bibfield  {journal} {\bibinfo  {journal} {Prog. Theor. Phys. Suppl.}\ }\textbf {\bibinfo {volume} {37}},\ \bibinfo {pages} {21} (\bibinfo {year} {1966})}\BibitemShut {NoStop}%
\bibitem [{\citenamefont {Kharzeev}(1996)}]{Kharzeev:1995ij}%
  \BibitemOpen
  \bibfield  {author} {\bibinfo {author} {\bibfnamefont {D.}~\bibnamefont {Kharzeev}},\ }\href {\doibase 10.3254/978-1-61499-215-8-105} {\bibfield  {journal} {\bibinfo  {journal} {Proc. Int. Sch. Phys. Fermi}\ }\textbf {\bibinfo {volume} {130}},\ \bibinfo {pages} {105} (\bibinfo {year} {1996})},\ \Eprint {http://arxiv.org/abs/nucl-th/9601029} {arXiv:nucl-th/9601029} \BibitemShut {NoStop}%
\bibitem [{\citenamefont {Shifman}\ \emph {et~al.}(1978)\citenamefont {Shifman}, \citenamefont {Vainshtein},\ and\ \citenamefont {Zakharov}}]{Shifman:1978zn}%
  \BibitemOpen
  \bibfield  {author} {\bibinfo {author} {\bibfnamefont {M.~A.}\ \bibnamefont {Shifman}}, \bibinfo {author} {\bibfnamefont {A.~I.}\ \bibnamefont {Vainshtein}}, \ and\ \bibinfo {author} {\bibfnamefont {V.~I.}\ \bibnamefont {Zakharov}},\ }\href {\doibase 10.1016/0370-2693(78)90481-1} {\bibfield  {journal} {\bibinfo  {journal} {Phys. Lett. B}\ }\textbf {\bibinfo {volume} {78}},\ \bibinfo {pages} {443} (\bibinfo {year} {1978})}\BibitemShut {NoStop}%
\bibitem [{\citenamefont {Ji}(1995)}]{Ji:1994av}%
  \BibitemOpen
  \bibfield  {author} {\bibinfo {author} {\bibfnamefont {X.-D.}\ \bibnamefont {Ji}},\ }\href {\doibase 10.1103/PhysRevLett.74.1071} {\bibfield  {journal} {\bibinfo  {journal} {Phys. Rev. Lett.}\ }\textbf {\bibinfo {volume} {74}},\ \bibinfo {pages} {1071} (\bibinfo {year} {1995})},\ \Eprint {http://arxiv.org/abs/hep-ph/9410274} {arXiv:hep-ph/9410274} \BibitemShut {NoStop}%
\bibitem [{\citenamefont {Aaij}\ \emph {et~al.}(2015)\citenamefont {Aaij} \emph {et~al.}}]{LHCb:2015yax}%
  \BibitemOpen
  \bibfield  {author} {\bibinfo {author} {\bibfnamefont {R.}~\bibnamefont {Aaij}} \emph {et~al.} (\bibinfo {collaboration} {LHCb}),\ }\href {\doibase 10.1103/PhysRevLett.115.072001} {\bibfield  {journal} {\bibinfo  {journal} {Phys. Rev. Lett.}\ }\textbf {\bibinfo {volume} {115}},\ \bibinfo {pages} {072001} (\bibinfo {year} {2015})},\ \Eprint {http://arxiv.org/abs/1507.03414} {arXiv:1507.03414 [hep-ex]} \BibitemShut {NoStop}%
\bibitem [{\citenamefont {Aaij}\ \emph {et~al.}(2019)\citenamefont {Aaij} \emph {et~al.}}]{LHCb:2019kea}%
  \BibitemOpen
  \bibfield  {author} {\bibinfo {author} {\bibfnamefont {R.}~\bibnamefont {Aaij}} \emph {et~al.} (\bibinfo {collaboration} {LHCb}),\ }\href {\doibase 10.1103/PhysRevLett.122.222001} {\bibfield  {journal} {\bibinfo  {journal} {Phys. Rev. Lett.}\ }\textbf {\bibinfo {volume} {122}},\ \bibinfo {pages} {222001} (\bibinfo {year} {2019})},\ \Eprint {http://arxiv.org/abs/1904.03947} {arXiv:1904.03947 [hep-ex]} \BibitemShut {NoStop}%
\bibitem [{\citenamefont {Sibirtsev}\ and\ \citenamefont {Voloshin}(2005)}]{Sibirtsev:2005ex}%
  \BibitemOpen
  \bibfield  {author} {\bibinfo {author} {\bibfnamefont {A.}~\bibnamefont {Sibirtsev}}\ and\ \bibinfo {author} {\bibfnamefont {M.~B.}\ \bibnamefont {Voloshin}},\ }\href {\doibase 10.1103/PhysRevD.71.076005} {\bibfield  {journal} {\bibinfo  {journal} {Phys. Rev. D}\ }\textbf {\bibinfo {volume} {71}},\ \bibinfo {pages} {076005} (\bibinfo {year} {2005})},\ \Eprint {http://arxiv.org/abs/hep-ph/0502068} {arXiv:hep-ph/0502068} \BibitemShut {NoStop}%
\bibitem [{\citenamefont {Brambilla}\ \emph {et~al.}(2014)\citenamefont {Brambilla} \emph {et~al.}}]{Brambilla:2014jmp}%
  \BibitemOpen
  \bibfield  {author} {\bibinfo {author} {\bibfnamefont {N.}~\bibnamefont {Brambilla}} \emph {et~al.},\ }\href {\doibase 10.1140/epjc/s10052-014-2981-5} {\bibfield  {journal} {\bibinfo  {journal} {Eur. Phys. J. C}\ }\textbf {\bibinfo {volume} {74}},\ \bibinfo {pages} {2981} (\bibinfo {year} {2014})},\ \Eprint {http://arxiv.org/abs/1404.3723} {arXiv:1404.3723 [hep-ph]} \BibitemShut {NoStop}%
\bibitem [{\citenamefont {Ali}\ \emph {et~al.}(2019)\citenamefont {Ali} \emph {et~al.}}]{GlueX:2019mkq}%
  \BibitemOpen
  \bibfield  {author} {\bibinfo {author} {\bibfnamefont {A.}~\bibnamefont {Ali}} \emph {et~al.} (\bibinfo {collaboration} {GlueX}),\ }\href {\doibase 10.1103/PhysRevLett.123.072001} {\bibfield  {journal} {\bibinfo  {journal} {Phys. Rev. Lett.}\ }\textbf {\bibinfo {volume} {123}},\ \bibinfo {pages} {072001} (\bibinfo {year} {2019})},\ \Eprint {http://arxiv.org/abs/1905.10811} {arXiv:1905.10811 [nucl-ex]} \BibitemShut {NoStop}%
\bibitem [{\citenamefont {Du}\ \emph {et~al.}(2020)\citenamefont {Du}, \citenamefont {Baru}, \citenamefont {Guo}, \citenamefont {Hanhart}, \citenamefont {Mei\ss{}ner}, \citenamefont {Nefediev},\ and\ \citenamefont {Strakovsky}}]{Du:2020bqj}%
  \BibitemOpen
  \bibfield  {author} {\bibinfo {author} {\bibfnamefont {M.-L.}\ \bibnamefont {Du}}, \bibinfo {author} {\bibfnamefont {V.}~\bibnamefont {Baru}}, \bibinfo {author} {\bibfnamefont {F.-K.}\ \bibnamefont {Guo}}, \bibinfo {author} {\bibfnamefont {C.}~\bibnamefont {Hanhart}}, \bibinfo {author} {\bibfnamefont {U.-G.}\ \bibnamefont {Mei\ss{}ner}}, \bibinfo {author} {\bibfnamefont {A.}~\bibnamefont {Nefediev}}, \ and\ \bibinfo {author} {\bibfnamefont {I.}~\bibnamefont {Strakovsky}},\ }\href {\doibase 10.1140/epjc/s10052-020-08620-5} {\bibfield  {journal} {\bibinfo  {journal} {Eur. Phys. J. C}\ }\textbf {\bibinfo {volume} {80}},\ \bibinfo {pages} {1053} (\bibinfo {year} {2020})},\ \Eprint {http://arxiv.org/abs/2009.08345} {arXiv:2009.08345 [hep-ph]} \BibitemShut {NoStop}%
\bibitem [{\citenamefont {Winney}\ \emph {et~al.}(2023)\citenamefont {Winney} \emph {et~al.}}]{JointPhysicsAnalysisCenter:2023qgg}%
  \BibitemOpen
  \bibfield  {author} {\bibinfo {author} {\bibfnamefont {D.}~\bibnamefont {Winney}} \emph {et~al.} (\bibinfo {collaboration} {Joint Physics Analysis Center}),\ }\href {\doibase 10.1103/PhysRevD.108.054018} {\bibfield  {journal} {\bibinfo  {journal} {Phys. Rev. D}\ }\textbf {\bibinfo {volume} {108}},\ \bibinfo {pages} {054018} (\bibinfo {year} {2023})},\ \Eprint {http://arxiv.org/abs/2305.01449} {arXiv:2305.01449 [hep-ph]} \BibitemShut {NoStop}%
\bibitem [{\citenamefont {Fabbietti}\ \emph {et~al.}(2021)\citenamefont {Fabbietti}, \citenamefont {Mantovani~Sarti},\ and\ \citenamefont {Vazquez~Doce}}]{Fabbietti:2020bfg}%
  \BibitemOpen
  \bibfield  {author} {\bibinfo {author} {\bibfnamefont {L.}~\bibnamefont {Fabbietti}}, \bibinfo {author} {\bibfnamefont {V.}~\bibnamefont {Mantovani~Sarti}}, \ and\ \bibinfo {author} {\bibfnamefont {O.}~\bibnamefont {Vazquez~Doce}},\ }\href {\doibase 10.1146/annurev-nucl-102419-034438} {\bibfield  {journal} {\bibinfo  {journal} {Ann. Rev. Nucl. Part. Sci.}\ }\textbf {\bibinfo {volume} {71}},\ \bibinfo {pages} {377} (\bibinfo {year} {2021})},\ \Eprint {http://arxiv.org/abs/2012.09806} {arXiv:2012.09806 [nucl-ex]} \BibitemShut {NoStop}%
\bibitem [{\citenamefont {Liu}\ \emph {et~al.}(2025)\citenamefont {Liu}, \citenamefont {Pan}, \citenamefont {Liu}, \citenamefont {Wu}, \citenamefont {Lu},\ and\ \citenamefont {Geng}}]{Liu:2024uxn}%
  \BibitemOpen
  \bibfield  {author} {\bibinfo {author} {\bibfnamefont {M.-Z.}\ \bibnamefont {Liu}}, \bibinfo {author} {\bibfnamefont {Y.-W.}\ \bibnamefont {Pan}}, \bibinfo {author} {\bibfnamefont {Z.-W.}\ \bibnamefont {Liu}}, \bibinfo {author} {\bibfnamefont {T.-W.}\ \bibnamefont {Wu}}, \bibinfo {author} {\bibfnamefont {J.-X.}\ \bibnamefont {Lu}}, \ and\ \bibinfo {author} {\bibfnamefont {L.-S.}\ \bibnamefont {Geng}},\ }\href {\doibase 10.1016/j.physrep.2024.12.001} {\bibfield  {journal} {\bibinfo  {journal} {Phys. Rept.}\ }\textbf {\bibinfo {volume} {1108}},\ \bibinfo {pages} {1} (\bibinfo {year} {2025})},\ \Eprint {http://arxiv.org/abs/2404.06399} {arXiv:2404.06399 [hep-ph]} \BibitemShut {NoStop}%
\bibitem [{\citenamefont {Adamczyk}\ \emph {et~al.}(2015{\natexlab{a}})\citenamefont {Adamczyk} \emph {et~al.}}]{STAR:2014dcy}%
  \BibitemOpen
  \bibfield  {author} {\bibinfo {author} {\bibfnamefont {L.}~\bibnamefont {Adamczyk}} \emph {et~al.} (\bibinfo {collaboration} {STAR}),\ }\href {\doibase 10.1103/PhysRevLett.114.022301} {\bibfield  {journal} {\bibinfo  {journal} {Phys. Rev. Lett.}\ }\textbf {\bibinfo {volume} {114}},\ \bibinfo {pages} {022301} (\bibinfo {year} {2015}{\natexlab{a}})},\ \Eprint {http://arxiv.org/abs/1408.4360} {arXiv:1408.4360 [nucl-ex]} \BibitemShut {NoStop}%
\bibitem [{\citenamefont {Adamczyk}\ \emph {et~al.}(2015{\natexlab{b}})\citenamefont {Adamczyk} \emph {et~al.}}]{STAR:2015kha}%
  \BibitemOpen
  \bibfield  {author} {\bibinfo {author} {\bibfnamefont {L.}~\bibnamefont {Adamczyk}} \emph {et~al.} (\bibinfo {collaboration} {STAR}),\ }\href {\doibase 10.1038/nature15724} {\bibfield  {journal} {\bibinfo  {journal} {Nature}\ }\textbf {\bibinfo {volume} {527}},\ \bibinfo {pages} {345} (\bibinfo {year} {2015}{\natexlab{b}})},\ \Eprint {http://arxiv.org/abs/1507.07158} {arXiv:1507.07158 [nucl-ex]} \BibitemShut {NoStop}%
\bibitem [{\citenamefont {Acharya}\ \emph {et~al.}(2020{\natexlab{a}})\citenamefont {Acharya} \emph {et~al.}}]{ALICE:2019gcn}%
  \BibitemOpen
  \bibfield  {author} {\bibinfo {author} {\bibfnamefont {S.}~\bibnamefont {Acharya}} \emph {et~al.} (\bibinfo {collaboration} {ALICE}),\ }\href {\doibase 10.1103/PhysRevLett.124.092301} {\bibfield  {journal} {\bibinfo  {journal} {Phys. Rev. Lett.}\ }\textbf {\bibinfo {volume} {124}},\ \bibinfo {pages} {092301} (\bibinfo {year} {2020}{\natexlab{a}})},\ \Eprint {http://arxiv.org/abs/1905.13470} {arXiv:1905.13470 [nucl-ex]} \BibitemShut {NoStop}%
\bibitem [{\citenamefont {Acharya}\ \emph {et~al.}(2019)\citenamefont {Acharya} \emph {et~al.}}]{ALICE:2019hdt}%
  \BibitemOpen
  \bibfield  {author} {\bibinfo {author} {\bibfnamefont {S.}~\bibnamefont {Acharya}} \emph {et~al.} (\bibinfo {collaboration} {ALICE}),\ }\href {\doibase 10.1103/PhysRevLett.123.112002} {\bibfield  {journal} {\bibinfo  {journal} {Phys. Rev. Lett.}\ }\textbf {\bibinfo {volume} {123}},\ \bibinfo {pages} {112002} (\bibinfo {year} {2019})},\ \Eprint {http://arxiv.org/abs/1904.12198} {arXiv:1904.12198 [nucl-ex]} \BibitemShut {NoStop}%
\bibitem [{\citenamefont {Collaboration}\ \emph {et~al.}(2020)\citenamefont {Collaboration} \emph {et~al.}}]{ALICE:2020mfd}%
  \BibitemOpen
  \bibfield  {author} {\bibinfo {author} {\bibfnamefont {A.}~\bibnamefont {Collaboration}} \emph {et~al.} (\bibinfo {collaboration} {ALICE}),\ }\href {\doibase 10.1038/s41586-020-3001-6} {\bibfield  {journal} {\bibinfo  {journal} {Nature}\ }\textbf {\bibinfo {volume} {588}},\ \bibinfo {pages} {232} (\bibinfo {year} {2020})},\ \bibinfo {note} {[Erratum: Nature 590, E13 (2021)]},\ \Eprint {http://arxiv.org/abs/2005.11495} {arXiv:2005.11495 [nucl-ex]} \BibitemShut {NoStop}%
\bibitem [{\citenamefont {Acharya}\ \emph {et~al.}(2021)\citenamefont {Acharya} \emph {et~al.}}]{ALICE:2021cpv}%
  \BibitemOpen
  \bibfield  {author} {\bibinfo {author} {\bibfnamefont {S.}~\bibnamefont {Acharya}} \emph {et~al.} (\bibinfo {collaboration} {ALICE}),\ }\href {\doibase 10.1103/PhysRevLett.127.172301} {\bibfield  {journal} {\bibinfo  {journal} {Phys. Rev. Lett.}\ }\textbf {\bibinfo {volume} {127}},\ \bibinfo {pages} {172301} (\bibinfo {year} {2021})},\ \Eprint {http://arxiv.org/abs/2105.05578} {arXiv:2105.05578 [nucl-ex]} \BibitemShut {NoStop}%
\bibitem [{\citenamefont {Si}\ \emph {et~al.}(2025)\citenamefont {Si} \emph {et~al.}}]{Si:2025eou}%
  \BibitemOpen
  \bibfield  {author} {\bibinfo {author} {\bibfnamefont {D.}~\bibnamefont {Si}} \emph {et~al.},\ }\href@noop {} {\  (\bibinfo {year} {2025})},\ \Eprint {http://arxiv.org/abs/2501.09576} {arXiv:2501.09576 [nucl-ex]} \BibitemShut {NoStop}%
\bibitem [{\citenamefont {Morita}\ \emph {et~al.}(2015)\citenamefont {Morita}, \citenamefont {Furumoto},\ and\ \citenamefont {Ohnishi}}]{Morita:2014kza}%
  \BibitemOpen
  \bibfield  {author} {\bibinfo {author} {\bibfnamefont {K.}~\bibnamefont {Morita}}, \bibinfo {author} {\bibfnamefont {T.}~\bibnamefont {Furumoto}}, \ and\ \bibinfo {author} {\bibfnamefont {A.}~\bibnamefont {Ohnishi}},\ }\href {\doibase 10.1103/PhysRevC.91.024916} {\bibfield  {journal} {\bibinfo  {journal} {Phys. Rev. C}\ }\textbf {\bibinfo {volume} {91}},\ \bibinfo {pages} {024916} (\bibinfo {year} {2015})},\ \Eprint {http://arxiv.org/abs/1408.6682} {arXiv:1408.6682 [nucl-th]} \BibitemShut {NoStop}%
\bibitem [{\citenamefont {Haidenbauer}(2019)}]{Haidenbauer:2018jvl}%
  \BibitemOpen
  \bibfield  {author} {\bibinfo {author} {\bibfnamefont {J.}~\bibnamefont {Haidenbauer}},\ }\href {\doibase 10.1016/j.nuclphysa.2018.10.090} {\bibfield  {journal} {\bibinfo  {journal} {Nucl. Phys. A}\ }\textbf {\bibinfo {volume} {981}},\ \bibinfo {pages} {1} (\bibinfo {year} {2019})},\ \Eprint {http://arxiv.org/abs/1808.05049} {arXiv:1808.05049 [hep-ph]} \BibitemShut {NoStop}%
\bibitem [{\citenamefont {Kamiya}\ \emph {et~al.}(2020)\citenamefont {Kamiya}, \citenamefont {Hyodo}, \citenamefont {Morita}, \citenamefont {Ohnishi},\ and\ \citenamefont {Weise}}]{Kamiya:2019uiw}%
  \BibitemOpen
  \bibfield  {author} {\bibinfo {author} {\bibfnamefont {Y.}~\bibnamefont {Kamiya}}, \bibinfo {author} {\bibfnamefont {T.}~\bibnamefont {Hyodo}}, \bibinfo {author} {\bibfnamefont {K.}~\bibnamefont {Morita}}, \bibinfo {author} {\bibfnamefont {A.}~\bibnamefont {Ohnishi}}, \ and\ \bibinfo {author} {\bibfnamefont {W.}~\bibnamefont {Weise}},\ }\href {\doibase 10.1103/PhysRevLett.124.132501} {\bibfield  {journal} {\bibinfo  {journal} {Phys. Rev. Lett.}\ }\textbf {\bibinfo {volume} {124}},\ \bibinfo {pages} {132501} (\bibinfo {year} {2020})},\ \Eprint {http://arxiv.org/abs/1911.01041} {arXiv:1911.01041 [nucl-th]} \BibitemShut {NoStop}%
\bibitem [{\citenamefont {Liu}\ \emph {et~al.}(2023{\natexlab{a}})\citenamefont {Liu}, \citenamefont {Li},\ and\ \citenamefont {Geng}}]{Liu:2022nec}%
  \BibitemOpen
  \bibfield  {author} {\bibinfo {author} {\bibfnamefont {Z.-W.}\ \bibnamefont {Liu}}, \bibinfo {author} {\bibfnamefont {K.-W.}\ \bibnamefont {Li}}, \ and\ \bibinfo {author} {\bibfnamefont {L.-S.}\ \bibnamefont {Geng}},\ }\href {\doibase 10.1088/1674-1137/ac988a} {\bibfield  {journal} {\bibinfo  {journal} {Chin. Phys. C}\ }\textbf {\bibinfo {volume} {47}},\ \bibinfo {pages} {024108} (\bibinfo {year} {2023}{\natexlab{a}})},\ \Eprint {http://arxiv.org/abs/2201.04997} {arXiv:2201.04997 [hep-ph]} \BibitemShut {NoStop}%
\bibitem [{\citenamefont {Molina}\ \emph {et~al.}(2024)\citenamefont {Molina}, \citenamefont {Liu}, \citenamefont {Geng},\ and\ \citenamefont {Oset}}]{Molina:2023oeu}%
  \BibitemOpen
  \bibfield  {author} {\bibinfo {author} {\bibfnamefont {R.}~\bibnamefont {Molina}}, \bibinfo {author} {\bibfnamefont {Z.-W.}\ \bibnamefont {Liu}}, \bibinfo {author} {\bibfnamefont {L.-S.}\ \bibnamefont {Geng}}, \ and\ \bibinfo {author} {\bibfnamefont {E.}~\bibnamefont {Oset}},\ }\href {\doibase 10.1140/epjc/s10052-024-12694-w} {\bibfield  {journal} {\bibinfo  {journal} {Eur. Phys. J. C}\ }\textbf {\bibinfo {volume} {84}},\ \bibinfo {pages} {328} (\bibinfo {year} {2024})},\ \Eprint {http://arxiv.org/abs/2312.11993} {arXiv:2312.11993 [hep-ph]} \BibitemShut {NoStop}%
\bibitem [{\citenamefont {Yan}\ \emph {et~al.}(2025)\citenamefont {Yan}, \citenamefont {Huang}, \citenamefont {Yang}, \citenamefont {Huang},\ and\ \citenamefont {Ping}}]{Yan:2024aap}%
  \BibitemOpen
  \bibfield  {author} {\bibinfo {author} {\bibfnamefont {Y.}~\bibnamefont {Yan}}, \bibinfo {author} {\bibfnamefont {Q.}~\bibnamefont {Huang}}, \bibinfo {author} {\bibfnamefont {Y.}~\bibnamefont {Yang}}, \bibinfo {author} {\bibfnamefont {H.}~\bibnamefont {Huang}}, \ and\ \bibinfo {author} {\bibfnamefont {J.}~\bibnamefont {Ping}},\ }\href {\doibase 10.1007/s11433-024-2580-9} {\bibfield  {journal} {\bibinfo  {journal} {Sci. China Phys. Mech. Astron.}\ }\textbf {\bibinfo {volume} {68}},\ \bibinfo {pages} {232012} (\bibinfo {year} {2025})},\ \Eprint {http://arxiv.org/abs/2408.15493} {arXiv:2408.15493 [hep-ph]} \BibitemShut {NoStop}%
\bibitem [{\citenamefont {Achenbach}\ \emph {et~al.}(2024)\citenamefont {Achenbach} \emph {et~al.}}]{Achenbach:2024wgy}%
  \BibitemOpen
  \bibfield  {author} {\bibinfo {author} {\bibfnamefont {P.}~\bibnamefont {Achenbach}} \emph {et~al.},\ }\href@noop {} {\  (\bibinfo {year} {2024})},\ \Eprint {http://arxiv.org/abs/2409.00366} {arXiv:2409.00366 [nucl-ex]} \BibitemShut {NoStop}%
\bibitem [{\citenamefont {Li}\ \emph {et~al.}(2024{\natexlab{a}})\citenamefont {Li}, \citenamefont {Xiao}, \citenamefont {Liang}, \citenamefont {Wu}, \citenamefont {Wang},\ and\ \citenamefont {Oset}}]{Li:2024tvo}%
  \BibitemOpen
  \bibfield  {author} {\bibinfo {author} {\bibfnamefont {H.-P.}\ \bibnamefont {Li}}, \bibinfo {author} {\bibfnamefont {C.-W.}\ \bibnamefont {Xiao}}, \bibinfo {author} {\bibfnamefont {W.-H.}\ \bibnamefont {Liang}}, \bibinfo {author} {\bibfnamefont {J.-J.}\ \bibnamefont {Wu}}, \bibinfo {author} {\bibfnamefont {E.}~\bibnamefont {Wang}}, \ and\ \bibinfo {author} {\bibfnamefont {E.}~\bibnamefont {Oset}},\ }\href {\doibase 10.1103/PhysRevD.110.114018} {\bibfield  {journal} {\bibinfo  {journal} {Phys. Rev. D}\ }\textbf {\bibinfo {volume} {110}},\ \bibinfo {pages} {114018} (\bibinfo {year} {2024}{\natexlab{a}})},\ \Eprint {http://arxiv.org/abs/2409.05787} {arXiv:2409.05787 [hep-ph]} \BibitemShut {NoStop}%
\bibitem [{\citenamefont {Ge}\ \emph {et~al.}(2025)\citenamefont {Ge}, \citenamefont {Liu}, \citenamefont {Lu},\ and\ \citenamefont {Geng}}]{Ge:2025put}%
  \BibitemOpen
  \bibfield  {author} {\bibinfo {author} {\bibfnamefont {D.-L.}\ \bibnamefont {Ge}}, \bibinfo {author} {\bibfnamefont {Z.-W.}\ \bibnamefont {Liu}}, \bibinfo {author} {\bibfnamefont {J.-X.}\ \bibnamefont {Lu}}, \ and\ \bibinfo {author} {\bibfnamefont {L.-S.}\ \bibnamefont {Geng}},\ }\href@noop {} {\  (\bibinfo {year} {2025})},\ \Eprint {http://arxiv.org/abs/2502.18872} {arXiv:2502.18872 [nucl-th]} \BibitemShut {NoStop}%
\bibitem [{\citenamefont {Ikeno}(2025)}]{Ikeno:2025kwe}%
  \BibitemOpen
  \bibfield  {author} {\bibinfo {author} {\bibfnamefont {N.}~\bibnamefont {Ikeno}},\ }\href@noop {} {\  (\bibinfo {year} {2025})},\ \Eprint {http://arxiv.org/abs/2502.20020} {arXiv:2502.20020 [nucl-th]} \BibitemShut {NoStop}%
\bibitem [{\citenamefont {Liu}\ and\ \citenamefont {Xie}(2025)}]{Liu:2025eqw}%
  \BibitemOpen
  \bibfield  {author} {\bibinfo {author} {\bibfnamefont {S.-W.}\ \bibnamefont {Liu}}\ and\ \bibinfo {author} {\bibfnamefont {J.-J.}\ \bibnamefont {Xie}},\ }\href@noop {} {\  (\bibinfo {year} {2025})},\ \Eprint {http://arxiv.org/abs/2503.22453} {arXiv:2503.22453 [hep-ph]} \BibitemShut {NoStop}%
\bibitem [{\citenamefont {Kamiya}\ \emph {et~al.}(2022)\citenamefont {Kamiya}, \citenamefont {Hyodo},\ and\ \citenamefont {Ohnishi}}]{Kamiya:2022thy}%
  \BibitemOpen
  \bibfield  {author} {\bibinfo {author} {\bibfnamefont {Y.}~\bibnamefont {Kamiya}}, \bibinfo {author} {\bibfnamefont {T.}~\bibnamefont {Hyodo}}, \ and\ \bibinfo {author} {\bibfnamefont {A.}~\bibnamefont {Ohnishi}},\ }\href {\doibase 10.1140/epja/s10050-022-00782-y} {\bibfield  {journal} {\bibinfo  {journal} {Eur. Phys. J. A}\ }\textbf {\bibinfo {volume} {58}},\ \bibinfo {pages} {131} (\bibinfo {year} {2022})},\ \Eprint {http://arxiv.org/abs/2203.13814} {arXiv:2203.13814 [hep-ph]} \BibitemShut {NoStop}%
\bibitem [{\citenamefont {Liu}\ \emph {et~al.}(2023{\natexlab{b}})\citenamefont {Liu}, \citenamefont {Lu},\ and\ \citenamefont {Geng}}]{Liu:2023uly}%
  \BibitemOpen
  \bibfield  {author} {\bibinfo {author} {\bibfnamefont {Z.-W.}\ \bibnamefont {Liu}}, \bibinfo {author} {\bibfnamefont {J.-X.}\ \bibnamefont {Lu}}, \ and\ \bibinfo {author} {\bibfnamefont {L.-S.}\ \bibnamefont {Geng}},\ }\href {\doibase 10.1103/PhysRevD.107.074019} {\bibfield  {journal} {\bibinfo  {journal} {Phys. Rev. D}\ }\textbf {\bibinfo {volume} {107}},\ \bibinfo {pages} {074019} (\bibinfo {year} {2023}{\natexlab{b}})},\ \Eprint {http://arxiv.org/abs/2302.01046} {arXiv:2302.01046 [hep-ph]} \BibitemShut {NoStop}%
\bibitem [{\citenamefont {Liu}\ \emph {et~al.}(2023{\natexlab{c}})\citenamefont {Liu}, \citenamefont {Lu}, \citenamefont {Liu},\ and\ \citenamefont {Geng}}]{Liu:2023wfo}%
  \BibitemOpen
  \bibfield  {author} {\bibinfo {author} {\bibfnamefont {Z.-W.}\ \bibnamefont {Liu}}, \bibinfo {author} {\bibfnamefont {J.-X.}\ \bibnamefont {Lu}}, \bibinfo {author} {\bibfnamefont {M.-Z.}\ \bibnamefont {Liu}}, \ and\ \bibinfo {author} {\bibfnamefont {L.-S.}\ \bibnamefont {Geng}},\ }\href {\doibase 10.1103/PhysRevD.108.L031503} {\bibfield  {journal} {\bibinfo  {journal} {Phys. Rev. D}\ }\textbf {\bibinfo {volume} {108}},\ \bibinfo {pages} {L031503} (\bibinfo {year} {2023}{\natexlab{c}})},\ \Eprint {http://arxiv.org/abs/2305.19048} {arXiv:2305.19048 [hep-ph]} \BibitemShut {NoStop}%
\bibitem [{\citenamefont {Vidana}\ \emph {et~al.}(2023)\citenamefont {Vidana}, \citenamefont {Feijoo}, \citenamefont {Albaladejo}, \citenamefont {Nieves},\ and\ \citenamefont {Oset}}]{Vidana:2023olz}%
  \BibitemOpen
  \bibfield  {author} {\bibinfo {author} {\bibfnamefont {I.}~\bibnamefont {Vidana}}, \bibinfo {author} {\bibfnamefont {A.}~\bibnamefont {Feijoo}}, \bibinfo {author} {\bibfnamefont {M.}~\bibnamefont {Albaladejo}}, \bibinfo {author} {\bibfnamefont {J.}~\bibnamefont {Nieves}}, \ and\ \bibinfo {author} {\bibfnamefont {E.}~\bibnamefont {Oset}},\ }\href {\doibase 10.1016/j.physletb.2023.138201} {\bibfield  {journal} {\bibinfo  {journal} {Phys. Lett. B}\ }\textbf {\bibinfo {volume} {846}},\ \bibinfo {pages} {138201} (\bibinfo {year} {2023})},\ \Eprint {http://arxiv.org/abs/2303.06079} {arXiv:2303.06079 [hep-ph]} \BibitemShut {NoStop}%
\bibitem [{\citenamefont {Ikeno}\ \emph {et~al.}(2023)\citenamefont {Ikeno}, \citenamefont {Toledo},\ and\ \citenamefont {Oset}}]{Ikeno:2023ojl}%
  \BibitemOpen
  \bibfield  {author} {\bibinfo {author} {\bibfnamefont {N.}~\bibnamefont {Ikeno}}, \bibinfo {author} {\bibfnamefont {G.}~\bibnamefont {Toledo}}, \ and\ \bibinfo {author} {\bibfnamefont {E.}~\bibnamefont {Oset}},\ }\href {\doibase 10.1016/j.physletb.2023.138281} {\bibfield  {journal} {\bibinfo  {journal} {Phys. Lett. B}\ }\textbf {\bibinfo {volume} {847}},\ \bibinfo {pages} {138281} (\bibinfo {year} {2023})},\ \Eprint {http://arxiv.org/abs/2305.16431} {arXiv:2305.16431 [hep-ph]} \BibitemShut {NoStop}%
\bibitem [{\citenamefont {Torres-Rincon}\ \emph {et~al.}(2023)\citenamefont {Torres-Rincon}, \citenamefont {Ramos},\ and\ \citenamefont {Tolos}}]{Torres-Rincon:2023qll}%
  \BibitemOpen
  \bibfield  {author} {\bibinfo {author} {\bibfnamefont {J.~M.}\ \bibnamefont {Torres-Rincon}}, \bibinfo {author} {\bibfnamefont {A.}~\bibnamefont {Ramos}}, \ and\ \bibinfo {author} {\bibfnamefont {L.}~\bibnamefont {Tolos}},\ }\href {\doibase 10.1103/PhysRevD.108.096008} {\bibfield  {journal} {\bibinfo  {journal} {Phys. Rev. D}\ }\textbf {\bibinfo {volume} {108}},\ \bibinfo {pages} {096008} (\bibinfo {year} {2023})},\ \Eprint {http://arxiv.org/abs/2307.02102} {arXiv:2307.02102 [hep-ph]} \BibitemShut {NoStop}%
\bibitem [{\citenamefont {Feijoo}\ \emph {et~al.}(2024)\citenamefont {Feijoo}, \citenamefont {Dai}, \citenamefont {Abreu},\ and\ \citenamefont {Oset}}]{Feijoo:2023sfe}%
  \BibitemOpen
  \bibfield  {author} {\bibinfo {author} {\bibfnamefont {A.}~\bibnamefont {Feijoo}}, \bibinfo {author} {\bibfnamefont {L.~R.}\ \bibnamefont {Dai}}, \bibinfo {author} {\bibfnamefont {L.~M.}\ \bibnamefont {Abreu}}, \ and\ \bibinfo {author} {\bibfnamefont {E.}~\bibnamefont {Oset}},\ }\href {\doibase 10.1103/PhysRevD.109.016014} {\bibfield  {journal} {\bibinfo  {journal} {Phys. Rev. D}\ }\textbf {\bibinfo {volume} {109}},\ \bibinfo {pages} {016014} (\bibinfo {year} {2024})},\ \Eprint {http://arxiv.org/abs/2309.00444} {arXiv:2309.00444 [hep-ph]} \BibitemShut {NoStop}%
\bibitem [{\citenamefont {Khemchandani}\ \emph {et~al.}(2024)\citenamefont {Khemchandani}, \citenamefont {Abreu}, \citenamefont {Martinez~Torres},\ and\ \citenamefont {Navarra}}]{Khemchandani:2023xup}%
  \BibitemOpen
  \bibfield  {author} {\bibinfo {author} {\bibfnamefont {K.~P.}\ \bibnamefont {Khemchandani}}, \bibinfo {author} {\bibfnamefont {L.~M.}\ \bibnamefont {Abreu}}, \bibinfo {author} {\bibfnamefont {A.}~\bibnamefont {Martinez~Torres}}, \ and\ \bibinfo {author} {\bibfnamefont {F.~S.}\ \bibnamefont {Navarra}},\ }\href {\doibase 10.1103/PhysRevD.110.036008} {\bibfield  {journal} {\bibinfo  {journal} {Phys. Rev. D}\ }\textbf {\bibinfo {volume} {110}},\ \bibinfo {pages} {036008} (\bibinfo {year} {2024})},\ \Eprint {http://arxiv.org/abs/2312.11811} {arXiv:2312.11811 [hep-ph]} \BibitemShut {NoStop}%
\bibitem [{\citenamefont {Li}\ \emph {et~al.}(2024{\natexlab{b}})\citenamefont {Li}, \citenamefont {Yi}, \citenamefont {Xiao}, \citenamefont {Yao}, \citenamefont {Liang},\ and\ \citenamefont {Oset}}]{Li:2024tof}%
  \BibitemOpen
  \bibfield  {author} {\bibinfo {author} {\bibfnamefont {H.-P.}\ \bibnamefont {Li}}, \bibinfo {author} {\bibfnamefont {J.-Y.}\ \bibnamefont {Yi}}, \bibinfo {author} {\bibfnamefont {C.-W.}\ \bibnamefont {Xiao}}, \bibinfo {author} {\bibfnamefont {D.-L.}\ \bibnamefont {Yao}}, \bibinfo {author} {\bibfnamefont {W.-H.}\ \bibnamefont {Liang}}, \ and\ \bibinfo {author} {\bibfnamefont {E.}~\bibnamefont {Oset}},\ }\href {\doibase 10.1088/1674-1137/ad2dc2} {\bibfield  {journal} {\bibinfo  {journal} {Chin. Phys. C}\ }\textbf {\bibinfo {volume} {48}},\ \bibinfo {pages} {053107} (\bibinfo {year} {2024}{\natexlab{b}})},\ \Eprint {http://arxiv.org/abs/2401.14302} {arXiv:2401.14302 [hep-ph]} \BibitemShut {NoStop}%
\bibitem [{\citenamefont {Albaladejo}\ \emph {et~al.}(2024)\citenamefont {Albaladejo}, \citenamefont {Feijoo}, \citenamefont {Nieves}, \citenamefont {Oset},\ and\ \citenamefont {Vida\~na}}]{Albaladejo:2024lam}%
  \BibitemOpen
  \bibfield  {author} {\bibinfo {author} {\bibfnamefont {M.}~\bibnamefont {Albaladejo}}, \bibinfo {author} {\bibfnamefont {A.}~\bibnamefont {Feijoo}}, \bibinfo {author} {\bibfnamefont {J.}~\bibnamefont {Nieves}}, \bibinfo {author} {\bibfnamefont {E.}~\bibnamefont {Oset}}, \ and\ \bibinfo {author} {\bibfnamefont {I.}~\bibnamefont {Vida\~na}},\ }\href {\doibase 10.1103/PhysRevD.110.114052} {\bibfield  {journal} {\bibinfo  {journal} {Phys. Rev. D}\ }\textbf {\bibinfo {volume} {110}},\ \bibinfo {pages} {114052} (\bibinfo {year} {2024})},\ \Eprint {http://arxiv.org/abs/2410.08880} {arXiv:2410.08880 [hep-ph]} \BibitemShut {NoStop}%
\bibitem [{\citenamefont {Liu}\ \emph {et~al.}(2024)\citenamefont {Liu}, \citenamefont {Lu}, \citenamefont {Liu},\ and\ \citenamefont {Geng}}]{Liu:2024nac}%
  \BibitemOpen
  \bibfield  {author} {\bibinfo {author} {\bibfnamefont {Z.-W.}\ \bibnamefont {Liu}}, \bibinfo {author} {\bibfnamefont {J.-X.}\ \bibnamefont {Lu}}, \bibinfo {author} {\bibfnamefont {M.-Z.}\ \bibnamefont {Liu}}, \ and\ \bibinfo {author} {\bibfnamefont {L.-S.}\ \bibnamefont {Geng}},\ }\href@noop {} {\  (\bibinfo {year} {2024})},\ \Eprint {http://arxiv.org/abs/2404.18607} {arXiv:2404.18607 [hep-ph]} \BibitemShut {NoStop}%
\bibitem [{\citenamefont {Geng}\ \emph {et~al.}(2025)\citenamefont {Geng}, \citenamefont {Liu},\ and\ \citenamefont {Lu}}]{Geng:2025ruq}%
  \BibitemOpen
  \bibfield  {author} {\bibinfo {author} {\bibfnamefont {L.-S.}\ \bibnamefont {Geng}}, \bibinfo {author} {\bibfnamefont {Z.-W.}\ \bibnamefont {Liu}}, \ and\ \bibinfo {author} {\bibfnamefont {J.-X.}\ \bibnamefont {Lu}},\ }\href {\doibase 10.22323/1.465.0044} {\bibfield  {journal} {\bibinfo  {journal} {PoS}\ }\textbf {\bibinfo {volume} {QNP2024}},\ \bibinfo {pages} {044} (\bibinfo {year} {2025})}\BibitemShut {NoStop}%
\bibitem [{\citenamefont {Acharya}\ \emph {et~al.}(2022{\natexlab{a}})\citenamefont {Acharya} \emph {et~al.}}]{ALICE:2022enj}%
  \BibitemOpen
  \bibfield  {author} {\bibinfo {author} {\bibfnamefont {S.}~\bibnamefont {Acharya}} \emph {et~al.} (\bibinfo {collaboration} {ALICE}),\ }\href {\doibase 10.1103/PhysRevD.106.052010} {\bibfield  {journal} {\bibinfo  {journal} {Phys. Rev. D}\ }\textbf {\bibinfo {volume} {106}},\ \bibinfo {pages} {052010} (\bibinfo {year} {2022}{\natexlab{a}})},\ \Eprint {http://arxiv.org/abs/2201.05352} {arXiv:2201.05352 [nucl-ex]} \BibitemShut {NoStop}%
\bibitem [{\citenamefont {Acharya}\ \emph {et~al.}(2024{\natexlab{a}})\citenamefont {Acharya} \emph {et~al.}}]{ALICE:2024bhk}%
  \BibitemOpen
  \bibfield  {author} {\bibinfo {author} {\bibfnamefont {S.}~\bibnamefont {Acharya}} \emph {et~al.} (\bibinfo {collaboration} {ALICE}),\ }\href {\doibase 10.1103/PhysRevD.110.032004} {\bibfield  {journal} {\bibinfo  {journal} {Phys. Rev. D}\ }\textbf {\bibinfo {volume} {110}},\ \bibinfo {pages} {032004} (\bibinfo {year} {2024}{\natexlab{a}})},\ \Eprint {http://arxiv.org/abs/2401.13541} {arXiv:2401.13541 [nucl-ex]} \BibitemShut {NoStop}%
\bibitem [{\citenamefont {Appelquist}\ and\ \citenamefont {Fischler}(1978)}]{Appelquist:1978rt}%
  \BibitemOpen
  \bibfield  {author} {\bibinfo {author} {\bibfnamefont {T.}~\bibnamefont {Appelquist}}\ and\ \bibinfo {author} {\bibfnamefont {W.}~\bibnamefont {Fischler}},\ }\href {\doibase 10.1016/0370-2693(78)90587-7} {\bibfield  {journal} {\bibinfo  {journal} {Phys. Lett. B}\ }\textbf {\bibinfo {volume} {77}},\ \bibinfo {pages} {405} (\bibinfo {year} {1978})}\BibitemShut {NoStop}%
\bibitem [{\citenamefont {Brodsky}\ and\ \citenamefont {Miller}(1997)}]{Brodsky:1997gh}%
  \BibitemOpen
  \bibfield  {author} {\bibinfo {author} {\bibfnamefont {S.~J.}\ \bibnamefont {Brodsky}}\ and\ \bibinfo {author} {\bibfnamefont {G.~A.}\ \bibnamefont {Miller}},\ }\href {\doibase 10.1016/S0370-2693(97)01045-9} {\bibfield  {journal} {\bibinfo  {journal} {Phys. Lett. B}\ }\textbf {\bibinfo {volume} {412}},\ \bibinfo {pages} {125} (\bibinfo {year} {1997})},\ \Eprint {http://arxiv.org/abs/hep-ph/9707382} {arXiv:hep-ph/9707382} \BibitemShut {NoStop}%
\bibitem [{\citenamefont {Wu}\ \emph {et~al.}(2024)\citenamefont {Wu}, \citenamefont {Dong}, \citenamefont {Du}, \citenamefont {Guo},\ and\ \citenamefont {Zou}}]{Wu:2024xwy}%
  \BibitemOpen
  \bibfield  {author} {\bibinfo {author} {\bibfnamefont {B.}~\bibnamefont {Wu}}, \bibinfo {author} {\bibfnamefont {X.-K.}\ \bibnamefont {Dong}}, \bibinfo {author} {\bibfnamefont {M.-L.}\ \bibnamefont {Du}}, \bibinfo {author} {\bibfnamefont {F.-K.}\ \bibnamefont {Guo}}, \ and\ \bibinfo {author} {\bibfnamefont {B.-S.}\ \bibnamefont {Zou}},\ }\href@noop {} {\  (\bibinfo {year} {2024})},\ \Eprint {http://arxiv.org/abs/2410.19526} {arXiv:2410.19526 [hep-ph]} \BibitemShut {NoStop}%
\bibitem [{\citenamefont {Lipkin}\ and\ \citenamefont {Zou}(1996)}]{Lipkin:1996ny}%
  \BibitemOpen
  \bibfield  {author} {\bibinfo {author} {\bibfnamefont {H.~J.}\ \bibnamefont {Lipkin}}\ and\ \bibinfo {author} {\bibfnamefont {B.-s.}\ \bibnamefont {Zou}},\ }\href {\doibase 10.1103/PhysRevD.53.6693} {\bibfield  {journal} {\bibinfo  {journal} {Phys. Rev. D}\ }\textbf {\bibinfo {volume} {53}},\ \bibinfo {pages} {6693} (\bibinfo {year} {1996})}\BibitemShut {NoStop}%
\bibitem [{\citenamefont {Yokokawa}\ \emph {et~al.}(2006)\citenamefont {Yokokawa}, \citenamefont {Sasaki}, \citenamefont {Hatsuda},\ and\ \citenamefont {Hayashigaki}}]{Yokokawa:2006td}%
  \BibitemOpen
  \bibfield  {author} {\bibinfo {author} {\bibfnamefont {K.}~\bibnamefont {Yokokawa}}, \bibinfo {author} {\bibfnamefont {S.}~\bibnamefont {Sasaki}}, \bibinfo {author} {\bibfnamefont {T.}~\bibnamefont {Hatsuda}}, \ and\ \bibinfo {author} {\bibfnamefont {A.}~\bibnamefont {Hayashigaki}},\ }\href {\doibase 10.1103/PhysRevD.74.034504} {\bibfield  {journal} {\bibinfo  {journal} {Phys. Rev. D}\ }\textbf {\bibinfo {volume} {74}},\ \bibinfo {pages} {034504} (\bibinfo {year} {2006})},\ \Eprint {http://arxiv.org/abs/hep-lat/0605009} {arXiv:hep-lat/0605009} \BibitemShut {NoStop}%
\bibitem [{\citenamefont {Liu}\ \emph {et~al.}(2008)\citenamefont {Liu}, \citenamefont {Lin},\ and\ \citenamefont {Orginos}}]{Liu:2008rza}%
  \BibitemOpen
  \bibfield  {author} {\bibinfo {author} {\bibfnamefont {L.}~\bibnamefont {Liu}}, \bibinfo {author} {\bibfnamefont {H.-W.}\ \bibnamefont {Lin}}, \ and\ \bibinfo {author} {\bibfnamefont {K.}~\bibnamefont {Orginos}},\ }\href {\doibase 10.22323/1.066.0112} {\bibfield  {journal} {\bibinfo  {journal} {PoS}\ }\textbf {\bibinfo {volume} {LATTICE2008}},\ \bibinfo {pages} {112} (\bibinfo {year} {2008})},\ \Eprint {http://arxiv.org/abs/0810.5412} {arXiv:0810.5412 [hep-lat]} \BibitemShut {NoStop}%
\bibitem [{\citenamefont {Kawanai}\ and\ \citenamefont {Sasaki}(2010)}]{Kawanai:2010ev}%
  \BibitemOpen
  \bibfield  {author} {\bibinfo {author} {\bibfnamefont {T.}~\bibnamefont {Kawanai}}\ and\ \bibinfo {author} {\bibfnamefont {S.}~\bibnamefont {Sasaki}},\ }\href {\doibase 10.1103/PhysRevD.82.091501} {\bibfield  {journal} {\bibinfo  {journal} {Phys. Rev. D}\ }\textbf {\bibinfo {volume} {82}},\ \bibinfo {pages} {091501} (\bibinfo {year} {2010})},\ \Eprint {http://arxiv.org/abs/1009.3332} {arXiv:1009.3332 [hep-lat]} \BibitemShut {NoStop}%
\bibitem [{\citenamefont {Alberti}\ \emph {et~al.}(2017)\citenamefont {Alberti}, \citenamefont {Bali}, \citenamefont {Collins}, \citenamefont {Knechtli}, \citenamefont {Moir},\ and\ \citenamefont {S\"oldner}}]{Alberti:2016dru}%
  \BibitemOpen
  \bibfield  {author} {\bibinfo {author} {\bibfnamefont {M.}~\bibnamefont {Alberti}}, \bibinfo {author} {\bibfnamefont {G.~S.}\ \bibnamefont {Bali}}, \bibinfo {author} {\bibfnamefont {S.}~\bibnamefont {Collins}}, \bibinfo {author} {\bibfnamefont {F.}~\bibnamefont {Knechtli}}, \bibinfo {author} {\bibfnamefont {G.}~\bibnamefont {Moir}}, \ and\ \bibinfo {author} {\bibfnamefont {W.}~\bibnamefont {S\"oldner}},\ }\href {\doibase 10.1103/PhysRevD.95.074501} {\bibfield  {journal} {\bibinfo  {journal} {Phys. Rev. D}\ }\textbf {\bibinfo {volume} {95}},\ \bibinfo {pages} {074501} (\bibinfo {year} {2017})},\ \Eprint {http://arxiv.org/abs/1608.06537} {arXiv:1608.06537 [hep-lat]} \BibitemShut {NoStop}%
\bibitem [{\citenamefont {Sugiura}\ \emph {et~al.}(2019)\citenamefont {Sugiura}, \citenamefont {Ikeda},\ and\ \citenamefont {Ishii}}]{Sugiura:2019pye}%
  \BibitemOpen
  \bibfield  {author} {\bibinfo {author} {\bibfnamefont {T.}~\bibnamefont {Sugiura}}, \bibinfo {author} {\bibfnamefont {Y.}~\bibnamefont {Ikeda}}, \ and\ \bibinfo {author} {\bibfnamefont {N.}~\bibnamefont {Ishii}},\ }\href {\doibase 10.22323/1.334.0093} {\bibfield  {journal} {\bibinfo  {journal} {PoS}\ }\textbf {\bibinfo {volume} {LATTICE2018}},\ \bibinfo {pages} {093} (\bibinfo {year} {2019})},\ \Eprint {http://arxiv.org/abs/1905.02336} {arXiv:1905.02336 [nucl-th]} \BibitemShut {NoStop}%
\bibitem [{\citenamefont {Skerbis}\ and\ \citenamefont {Prelovsek}(2019)}]{Skerbis:2018lew}%
  \BibitemOpen
  \bibfield  {author} {\bibinfo {author} {\bibfnamefont {U.}~\bibnamefont {Skerbis}}\ and\ \bibinfo {author} {\bibfnamefont {S.}~\bibnamefont {Prelovsek}},\ }\href {\doibase 10.1103/PhysRevD.99.094505} {\bibfield  {journal} {\bibinfo  {journal} {Phys. Rev. D}\ }\textbf {\bibinfo {volume} {99}},\ \bibinfo {pages} {094505} (\bibinfo {year} {2019})},\ \Eprint {http://arxiv.org/abs/1811.02285} {arXiv:1811.02285 [hep-lat]} \BibitemShut {NoStop}%
\bibitem [{\citenamefont {Lyu}\ \emph {et~al.}(2025)\citenamefont {Lyu}, \citenamefont {Doi}, \citenamefont {Hatsuda},\ and\ \citenamefont {Sugiura}}]{Lyu:2024ttm}%
  \BibitemOpen
  \bibfield  {author} {\bibinfo {author} {\bibfnamefont {Y.}~\bibnamefont {Lyu}}, \bibinfo {author} {\bibfnamefont {T.}~\bibnamefont {Doi}}, \bibinfo {author} {\bibfnamefont {T.}~\bibnamefont {Hatsuda}}, \ and\ \bibinfo {author} {\bibfnamefont {T.}~\bibnamefont {Sugiura}},\ }\href {\doibase 10.1016/j.physletb.2024.139178} {\bibfield  {journal} {\bibinfo  {journal} {Phys. Lett. B}\ }\textbf {\bibinfo {volume} {860}},\ \bibinfo {pages} {139178} (\bibinfo {year} {2025})},\ \Eprint {http://arxiv.org/abs/2410.22755} {arXiv:2410.22755 [hep-lat]} \BibitemShut {NoStop}%
\bibitem [{\citenamefont {Hayashigaki}(1999)}]{Hayashigaki:1998ey}%
  \BibitemOpen
  \bibfield  {author} {\bibinfo {author} {\bibfnamefont {A.}~\bibnamefont {Hayashigaki}},\ }\href {\doibase 10.1143/PTP.101.923} {\bibfield  {journal} {\bibinfo  {journal} {Prog. Theor. Phys.}\ }\textbf {\bibinfo {volume} {101}},\ \bibinfo {pages} {923} (\bibinfo {year} {1999})},\ \Eprint {http://arxiv.org/abs/nucl-th/9811092} {arXiv:nucl-th/9811092} \BibitemShut {NoStop}%
\bibitem [{\citenamefont {Wen}\ \emph {et~al.}(2025)\citenamefont {Wen}, \citenamefont {Ma}, \citenamefont {Meng},\ and\ \citenamefont {Zhu}}]{Wen:2025wit}%
  \BibitemOpen
  \bibfield  {author} {\bibinfo {author} {\bibfnamefont {L.-Z.}\ \bibnamefont {Wen}}, \bibinfo {author} {\bibfnamefont {Y.}~\bibnamefont {Ma}}, \bibinfo {author} {\bibfnamefont {L.}~\bibnamefont {Meng}}, \ and\ \bibinfo {author} {\bibfnamefont {S.-L.}\ \bibnamefont {Zhu}},\ }\href@noop {} {\  (\bibinfo {year} {2025})},\ \Eprint {http://arxiv.org/abs/2503.11938} {arXiv:2503.11938 [hep-ph]} \BibitemShut {NoStop}%
\bibitem [{\citenamefont {Krein}\ and\ \citenamefont {Peixoto}(2020)}]{Krein:2020yor}%
  \BibitemOpen
  \bibfield  {author} {\bibinfo {author} {\bibfnamefont {G.}~\bibnamefont {Krein}}\ and\ \bibinfo {author} {\bibfnamefont {T.~C.}\ \bibnamefont {Peixoto}},\ }\href {\doibase 10.1007/s00601-020-01581-1} {\bibfield  {journal} {\bibinfo  {journal} {Few Body Syst.}\ }\textbf {\bibinfo {volume} {61}},\ \bibinfo {pages} {49} (\bibinfo {year} {2020})},\ \Eprint {http://arxiv.org/abs/2011.11615} {arXiv:2011.11615 [hep-ph]} \BibitemShut {NoStop}%
\bibitem [{\citenamefont {Krein}(2022)}]{Krein:2022fhf}%
  \BibitemOpen
  \bibfield  {author} {\bibinfo {author} {\bibfnamefont {G.~a.}\ \bibnamefont {Krein}},\ }\href {\doibase 10.1051/epjconf/202227404003} {\bibfield  {journal} {\bibinfo  {journal} {EPJ Web Conf.}\ }\textbf {\bibinfo {volume} {274}},\ \bibinfo {pages} {04003} (\bibinfo {year} {2022})}\BibitemShut {NoStop}%
\bibitem [{\citenamefont {Krein}(2023)}]{Krein:2023azg}%
  \BibitemOpen
  \bibfield  {author} {\bibinfo {author} {\bibfnamefont {G.~a.}\ \bibnamefont {Krein}},\ }\href {\doibase 10.1007/s00601-023-01829-6} {\bibfield  {journal} {\bibinfo  {journal} {Few Body Syst.}\ }\textbf {\bibinfo {volume} {64}},\ \bibinfo {pages} {42} (\bibinfo {year} {2023})}\BibitemShut {NoStop}%
\bibitem [{\citenamefont {Koonin}(1977)}]{Koonin:1977fh}%
  \BibitemOpen
  \bibfield  {author} {\bibinfo {author} {\bibfnamefont {S.~E.}\ \bibnamefont {Koonin}},\ }\href {\doibase 10.1016/0370-2693(77)90340-9} {\bibfield  {journal} {\bibinfo  {journal} {Phys. Lett. B}\ }\textbf {\bibinfo {volume} {70}},\ \bibinfo {pages} {43} (\bibinfo {year} {1977})}\BibitemShut {NoStop}%
\bibitem [{\citenamefont {Pratt}\ \emph {et~al.}(1990)\citenamefont {Pratt}, \citenamefont {Csorgo},\ and\ \citenamefont {Zimanyi}}]{Pratt:1990zq}%
  \BibitemOpen
  \bibfield  {author} {\bibinfo {author} {\bibfnamefont {S.}~\bibnamefont {Pratt}}, \bibinfo {author} {\bibfnamefont {T.}~\bibnamefont {Csorgo}}, \ and\ \bibinfo {author} {\bibfnamefont {J.}~\bibnamefont {Zimanyi}},\ }\href {\doibase 10.1103/PhysRevC.42.2646} {\bibfield  {journal} {\bibinfo  {journal} {Phys. Rev. C}\ }\textbf {\bibinfo {volume} {42}},\ \bibinfo {pages} {2646} (\bibinfo {year} {1990})}\BibitemShut {NoStop}%
\bibitem [{\citenamefont {Molina}\ and\ \citenamefont {Oset}(2025)}]{Molina:2025lzw}%
  \BibitemOpen
  \bibfield  {author} {\bibinfo {author} {\bibfnamefont {R.}~\bibnamefont {Molina}}\ and\ \bibinfo {author} {\bibfnamefont {E.}~\bibnamefont {Oset}},\ }\href@noop {} {\  (\bibinfo {year} {2025})},\ \Eprint {http://arxiv.org/abs/2506.03669} {arXiv:2506.03669 [hep-ph]} \BibitemShut {NoStop}%
\bibitem [{\citenamefont {Wiringa}\ \emph {et~al.}(1995)\citenamefont {Wiringa}, \citenamefont {Stoks},\ and\ \citenamefont {Schiavilla}}]{Wiringa:1994wb}%
  \BibitemOpen
  \bibfield  {author} {\bibinfo {author} {\bibfnamefont {R.~B.}\ \bibnamefont {Wiringa}}, \bibinfo {author} {\bibfnamefont {V.~G.~J.}\ \bibnamefont {Stoks}}, \ and\ \bibinfo {author} {\bibfnamefont {R.}~\bibnamefont {Schiavilla}},\ }\href {\doibase 10.1103/PhysRevC.51.38} {\bibfield  {journal} {\bibinfo  {journal} {Phys. Rev. C}\ }\textbf {\bibinfo {volume} {51}},\ \bibinfo {pages} {38} (\bibinfo {year} {1995})},\ \Eprint {http://arxiv.org/abs/nucl-th/9408016} {arXiv:nucl-th/9408016} \BibitemShut {NoStop}%
\bibitem [{\citenamefont {Machleidt}(2001)}]{Machleidt:2000ge}%
  \BibitemOpen
  \bibfield  {author} {\bibinfo {author} {\bibfnamefont {R.}~\bibnamefont {Machleidt}},\ }\href {\doibase 10.1103/PhysRevC.63.024001} {\bibfield  {journal} {\bibinfo  {journal} {Phys. Rev. C}\ }\textbf {\bibinfo {volume} {63}},\ \bibinfo {pages} {024001} (\bibinfo {year} {2001})},\ \Eprint {http://arxiv.org/abs/nucl-th/0006014} {arXiv:nucl-th/0006014} \BibitemShut {NoStop}%
\bibitem [{\citenamefont {Epelbaum}\ \emph {et~al.}(2015)\citenamefont {Epelbaum}, \citenamefont {Krebs},\ and\ \citenamefont {Mei\ss{}ner}}]{Epelbaum:2014sza}%
  \BibitemOpen
  \bibfield  {author} {\bibinfo {author} {\bibfnamefont {E.}~\bibnamefont {Epelbaum}}, \bibinfo {author} {\bibfnamefont {H.}~\bibnamefont {Krebs}}, \ and\ \bibinfo {author} {\bibfnamefont {U.~G.}\ \bibnamefont {Mei\ss{}ner}},\ }\href {\doibase 10.1103/PhysRevLett.115.122301} {\bibfield  {journal} {\bibinfo  {journal} {Phys. Rev. Lett.}\ }\textbf {\bibinfo {volume} {115}},\ \bibinfo {pages} {122301} (\bibinfo {year} {2015})},\ \Eprint {http://arxiv.org/abs/1412.4623} {arXiv:1412.4623 [nucl-th]} \BibitemShut {NoStop}%
\bibitem [{\citenamefont {Lu}\ \emph {et~al.}(2022)\citenamefont {Lu}, \citenamefont {Wang}, \citenamefont {Xiao}, \citenamefont {Geng}, \citenamefont {Meng},\ and\ \citenamefont {Ring}}]{Lu:2021gsb}%
  \BibitemOpen
  \bibfield  {author} {\bibinfo {author} {\bibfnamefont {J.-X.}\ \bibnamefont {Lu}}, \bibinfo {author} {\bibfnamefont {C.-X.}\ \bibnamefont {Wang}}, \bibinfo {author} {\bibfnamefont {Y.}~\bibnamefont {Xiao}}, \bibinfo {author} {\bibfnamefont {L.-S.}\ \bibnamefont {Geng}}, \bibinfo {author} {\bibfnamefont {J.}~\bibnamefont {Meng}}, \ and\ \bibinfo {author} {\bibfnamefont {P.}~\bibnamefont {Ring}},\ }\href {\doibase 10.1103/PhysRevLett.128.142002} {\bibfield  {journal} {\bibinfo  {journal} {Phys. Rev. Lett.}\ }\textbf {\bibinfo {volume} {128}},\ \bibinfo {pages} {142002} (\bibinfo {year} {2022})},\ \Eprint {http://arxiv.org/abs/2111.07766} {arXiv:2111.07766 [nucl-th]} \BibitemShut {NoStop}%
\bibitem [{\citenamefont {Lu}\ \emph {et~al.}(2025)\citenamefont {Lu}, \citenamefont {Xiao}, \citenamefont {Liu},\ and\ \citenamefont {Geng}}]{Lu:2025syk}%
  \BibitemOpen
  \bibfield  {author} {\bibinfo {author} {\bibfnamefont {J.-X.}\ \bibnamefont {Lu}}, \bibinfo {author} {\bibfnamefont {Y.}~\bibnamefont {Xiao}}, \bibinfo {author} {\bibfnamefont {Z.-W.}\ \bibnamefont {Liu}}, \ and\ \bibinfo {author} {\bibfnamefont {L.-S.}\ \bibnamefont {Geng}},\ }\href@noop {} {\  (\bibinfo {year} {2025})},\ \Eprint {http://arxiv.org/abs/2501.17185} {arXiv:2501.17185 [nucl-th]} \BibitemShut {NoStop}%
\bibitem [{\citenamefont {Epelbaum}\ \emph {et~al.}(2009)\citenamefont {Epelbaum}, \citenamefont {Hammer},\ and\ \citenamefont {Meissner}}]{Epelbaum:2008ga}%
  \BibitemOpen
  \bibfield  {author} {\bibinfo {author} {\bibfnamefont {E.}~\bibnamefont {Epelbaum}}, \bibinfo {author} {\bibfnamefont {H.-W.}\ \bibnamefont {Hammer}}, \ and\ \bibinfo {author} {\bibfnamefont {U.-G.}\ \bibnamefont {Meissner}},\ }\href {\doibase 10.1103/RevModPhys.81.1773} {\bibfield  {journal} {\bibinfo  {journal} {Rev. Mod. Phys.}\ }\textbf {\bibinfo {volume} {81}},\ \bibinfo {pages} {1773} (\bibinfo {year} {2009})},\ \Eprint {http://arxiv.org/abs/0811.1338} {arXiv:0811.1338 [nucl-th]} \BibitemShut {NoStop}%
\bibitem [{\citenamefont {Machleidt}\ and\ \citenamefont {Entem}(2011)}]{Machleidt:2011zz}%
  \BibitemOpen
  \bibfield  {author} {\bibinfo {author} {\bibfnamefont {R.}~\bibnamefont {Machleidt}}\ and\ \bibinfo {author} {\bibfnamefont {D.~R.}\ \bibnamefont {Entem}},\ }\href {\doibase 10.1016/j.physrep.2011.02.001} {\bibfield  {journal} {\bibinfo  {journal} {Phys. Rept.}\ }\textbf {\bibinfo {volume} {503}},\ \bibinfo {pages} {1} (\bibinfo {year} {2011})},\ \Eprint {http://arxiv.org/abs/1105.2919} {arXiv:1105.2919 [nucl-th]} \BibitemShut {NoStop}%
\bibitem [{\citenamefont {Hammer}\ \emph {et~al.}(2020)\citenamefont {Hammer}, \citenamefont {K\"onig},\ and\ \citenamefont {van Kolck}}]{Hammer:2019poc}%
  \BibitemOpen
  \bibfield  {author} {\bibinfo {author} {\bibfnamefont {H.~W.}\ \bibnamefont {Hammer}}, \bibinfo {author} {\bibfnamefont {S.}~\bibnamefont {K\"onig}}, \ and\ \bibinfo {author} {\bibfnamefont {U.}~\bibnamefont {van Kolck}},\ }\href {\doibase 10.1103/RevModPhys.92.025004} {\bibfield  {journal} {\bibinfo  {journal} {Rev. Mod. Phys.}\ }\textbf {\bibinfo {volume} {92}},\ \bibinfo {pages} {025004} (\bibinfo {year} {2020})},\ \Eprint {http://arxiv.org/abs/1906.12122} {arXiv:1906.12122 [nucl-th]} \BibitemShut {NoStop}%
\bibitem [{\citenamefont {Acharya}\ \emph {et~al.}(2020{\natexlab{b}})\citenamefont {Acharya} \emph {et~al.}}]{ALICE:2020ibs}%
  \BibitemOpen
  \bibfield  {author} {\bibinfo {author} {\bibfnamefont {S.}~\bibnamefont {Acharya}} \emph {et~al.} (\bibinfo {collaboration} {ALICE}),\ }\href {\doibase 10.1016/j.physletb.2020.135849} {\bibfield  {journal} {\bibinfo  {journal} {Phys. Lett. B}\ }\textbf {\bibinfo {volume} {811}},\ \bibinfo {pages} {135849} (\bibinfo {year} {2020}{\natexlab{b}})},\ \Eprint {http://arxiv.org/abs/2004.08018} {arXiv:2004.08018 [nucl-ex]} \BibitemShut {NoStop}%
\bibitem [{\citenamefont {Acharya}\ \emph {et~al.}(2025)\citenamefont {Acharya} \emph {et~al.}}]{ALICE:2023sjd}%
  \BibitemOpen
  \bibfield  {author} {\bibinfo {author} {\bibfnamefont {S.}~\bibnamefont {Acharya}} \emph {et~al.} (\bibinfo {collaboration} {ALICE}),\ }\href {\doibase 10.1140/epjc/s10052-025-13793-y} {\bibfield  {journal} {\bibinfo  {journal} {Eur. Phys. J. C}\ }\textbf {\bibinfo {volume} {85}},\ \bibinfo {pages} {198} (\bibinfo {year} {2025})},\ \Eprint {http://arxiv.org/abs/2311.14527} {arXiv:2311.14527 [hep-ph]} \BibitemShut {NoStop}%
\bibitem [{\citenamefont {Xu}\ \emph {et~al.}(2025)\citenamefont {Xu}, \citenamefont {Qin}, \citenamefont {Zou}, \citenamefont {Si}, \citenamefont {Xiao}, \citenamefont {Tian}, \citenamefont {Wang},\ and\ \citenamefont {Xiao}}]{Xu:2024dnd}%
  \BibitemOpen
  \bibfield  {author} {\bibinfo {author} {\bibfnamefont {J.}~\bibnamefont {Xu}}, \bibinfo {author} {\bibfnamefont {Z.}~\bibnamefont {Qin}}, \bibinfo {author} {\bibfnamefont {R.}~\bibnamefont {Zou}}, \bibinfo {author} {\bibfnamefont {D.}~\bibnamefont {Si}}, \bibinfo {author} {\bibfnamefont {S.}~\bibnamefont {Xiao}}, \bibinfo {author} {\bibfnamefont {B.}~\bibnamefont {Tian}}, \bibinfo {author} {\bibfnamefont {Y.}~\bibnamefont {Wang}}, \ and\ \bibinfo {author} {\bibfnamefont {Z.}~\bibnamefont {Xiao}},\ }\href {\doibase 10.1088/0256-307X/42/3/031401} {\bibfield  {journal} {\bibinfo  {journal} {Chin. Phys. Lett.}\ }\textbf {\bibinfo {volume} {42}},\ \bibinfo {pages} {031401} (\bibinfo {year} {2025})},\ \Eprint {http://arxiv.org/abs/2411.08718} {arXiv:2411.08718 [nucl-th]} \BibitemShut {NoStop}%
\bibitem [{\citenamefont {Wang}\ and\ \citenamefont {Zhao}(2024)}]{Wang:2024bpl}%
  \BibitemOpen
  \bibfield  {author} {\bibinfo {author} {\bibfnamefont {L.}~\bibnamefont {Wang}}\ and\ \bibinfo {author} {\bibfnamefont {J.}~\bibnamefont {Zhao}},\ }\href@noop {} {\  (\bibinfo {year} {2024})},\ \Eprint {http://arxiv.org/abs/2411.16343} {arXiv:2411.16343 [nucl-th]} \BibitemShut {NoStop}%
\bibitem [{\citenamefont {Feijoo}\ \emph {et~al.}(2025)\citenamefont {Feijoo}, \citenamefont {Korwieser},\ and\ \citenamefont {Fabbietti}}]{Feijoo:2024bvn}%
  \BibitemOpen
  \bibfield  {author} {\bibinfo {author} {\bibfnamefont {A.}~\bibnamefont {Feijoo}}, \bibinfo {author} {\bibfnamefont {M.}~\bibnamefont {Korwieser}}, \ and\ \bibinfo {author} {\bibfnamefont {L.}~\bibnamefont {Fabbietti}},\ }\href {\doibase 10.1103/PhysRevD.111.014009} {\bibfield  {journal} {\bibinfo  {journal} {Phys. Rev. D}\ }\textbf {\bibinfo {volume} {111}},\ \bibinfo {pages} {014009} (\bibinfo {year} {2025})},\ \Eprint {http://arxiv.org/abs/2407.01128} {arXiv:2407.01128 [hep-ph]} \BibitemShut {NoStop}%
\bibitem [{\citenamefont {Chizzali}\ \emph {et~al.}(2024)\citenamefont {Chizzali}, \citenamefont {Kamiya}, \citenamefont {Del~Grande}, \citenamefont {Doi}, \citenamefont {Fabbietti}, \citenamefont {Hatsuda},\ and\ \citenamefont {Lyu}}]{Chizzali:2022pjd}%
  \BibitemOpen
  \bibfield  {author} {\bibinfo {author} {\bibfnamefont {E.}~\bibnamefont {Chizzali}}, \bibinfo {author} {\bibfnamefont {Y.}~\bibnamefont {Kamiya}}, \bibinfo {author} {\bibfnamefont {R.}~\bibnamefont {Del~Grande}}, \bibinfo {author} {\bibfnamefont {T.}~\bibnamefont {Doi}}, \bibinfo {author} {\bibfnamefont {L.}~\bibnamefont {Fabbietti}}, \bibinfo {author} {\bibfnamefont {T.}~\bibnamefont {Hatsuda}}, \ and\ \bibinfo {author} {\bibfnamefont {Y.}~\bibnamefont {Lyu}},\ }\href {\doibase 10.1016/j.physletb.2023.138358} {\bibfield  {journal} {\bibinfo  {journal} {Phys. Lett. B}\ }\textbf {\bibinfo {volume} {848}},\ \bibinfo {pages} {138358} (\bibinfo {year} {2024})},\ \Eprint {http://arxiv.org/abs/2212.12690} {arXiv:2212.12690 [nucl-ex]} \BibitemShut {NoStop}%
\bibitem [{\citenamefont {Lyu}\ \emph {et~al.}(2022)\citenamefont {Lyu}, \citenamefont {Doi}, \citenamefont {Hatsuda}, \citenamefont {Ikeda}, \citenamefont {Meng}, \citenamefont {Sasaki},\ and\ \citenamefont {Sugiura}}]{Lyu:2022imf}%
  \BibitemOpen
  \bibfield  {author} {\bibinfo {author} {\bibfnamefont {Y.}~\bibnamefont {Lyu}}, \bibinfo {author} {\bibfnamefont {T.}~\bibnamefont {Doi}}, \bibinfo {author} {\bibfnamefont {T.}~\bibnamefont {Hatsuda}}, \bibinfo {author} {\bibfnamefont {Y.}~\bibnamefont {Ikeda}}, \bibinfo {author} {\bibfnamefont {J.}~\bibnamefont {Meng}}, \bibinfo {author} {\bibfnamefont {K.}~\bibnamefont {Sasaki}}, \ and\ \bibinfo {author} {\bibfnamefont {T.}~\bibnamefont {Sugiura}},\ }\href {\doibase 10.1103/PhysRevD.106.074507} {\bibfield  {journal} {\bibinfo  {journal} {Phys. Rev. D}\ }\textbf {\bibinfo {volume} {106}},\ \bibinfo {pages} {074507} (\bibinfo {year} {2022})},\ \Eprint {http://arxiv.org/abs/2205.10544} {arXiv:2205.10544 [hep-lat]} \BibitemShut {NoStop}%
\bibitem [{\citenamefont {Eides}\ \emph {et~al.}(2018)\citenamefont {Eides}, \citenamefont {Petrov},\ and\ \citenamefont {Polyakov}}]{Eides:2017xnt}%
  \BibitemOpen
  \bibfield  {author} {\bibinfo {author} {\bibfnamefont {M.~I.}\ \bibnamefont {Eides}}, \bibinfo {author} {\bibfnamefont {V.~Y.}\ \bibnamefont {Petrov}}, \ and\ \bibinfo {author} {\bibfnamefont {M.~V.}\ \bibnamefont {Polyakov}},\ }\href {\doibase 10.1140/epjc/s10052-018-5530-9} {\bibfield  {journal} {\bibinfo  {journal} {Eur. Phys. J. C}\ }\textbf {\bibinfo {volume} {78}},\ \bibinfo {pages} {36} (\bibinfo {year} {2018})},\ \Eprint {http://arxiv.org/abs/1709.09523} {arXiv:1709.09523 [hep-ph]} \BibitemShut {NoStop}%
\bibitem [{\citenamefont {Lednicky}\ and\ \citenamefont {Lyuboshits}(1981)}]{Lednicky:1981su}%
  \BibitemOpen
  \bibfield  {author} {\bibinfo {author} {\bibfnamefont {R.}~\bibnamefont {Lednicky}}\ and\ \bibinfo {author} {\bibfnamefont {V.~L.}\ \bibnamefont {Lyuboshits}},\ }\href@noop {} {\bibfield  {journal} {\bibinfo  {journal} {Yad. Fiz.}\ }\textbf {\bibinfo {volume} {35}},\ \bibinfo {pages} {1316} (\bibinfo {year} {1981})}\BibitemShut {NoStop}%
\bibitem [{\citenamefont {Cho}\ \emph {et~al.}(2017)\citenamefont {Cho} \emph {et~al.}}]{ExHIC:2017smd}%
  \BibitemOpen
  \bibfield  {author} {\bibinfo {author} {\bibfnamefont {S.}~\bibnamefont {Cho}} \emph {et~al.} (\bibinfo {collaboration} {ExHIC}),\ }\href {\doibase 10.1016/j.ppnp.2017.02.002} {\bibfield  {journal} {\bibinfo  {journal} {Prog. Part. Nucl. Phys.}\ }\textbf {\bibinfo {volume} {95}},\ \bibinfo {pages} {279} (\bibinfo {year} {2017})},\ \Eprint {http://arxiv.org/abs/1702.00486} {arXiv:1702.00486 [nucl-th]} \BibitemShut {NoStop}%
\bibitem [{\citenamefont {Aaij}\ \emph {et~al.}(2020)\citenamefont {Aaij} \emph {et~al.}}]{LHCb:2020bwg}%
  \BibitemOpen
  \bibfield  {author} {\bibinfo {author} {\bibfnamefont {R.}~\bibnamefont {Aaij}} \emph {et~al.} (\bibinfo {collaboration} {LHCb}),\ }\href {\doibase 10.1016/j.scib.2020.08.032} {\bibfield  {journal} {\bibinfo  {journal} {Sci. Bull.}\ }\textbf {\bibinfo {volume} {65}},\ \bibinfo {pages} {1983} (\bibinfo {year} {2020})},\ \Eprint {http://arxiv.org/abs/2006.16957} {arXiv:2006.16957 [hep-ex]} \BibitemShut {NoStop}%
\bibitem [{\citenamefont {Acharya}\ \emph {et~al.}(2024{\natexlab{b}})\citenamefont {Acharya} \emph {et~al.}}]{ALICE:2023hou}%
  \BibitemOpen
  \bibfield  {author} {\bibinfo {author} {\bibfnamefont {S.}~\bibnamefont {Acharya}} \emph {et~al.} (\bibinfo {collaboration} {ALICE}),\ }\href {\doibase 10.1007/JHEP02(2024)066} {\bibfield  {journal} {\bibinfo  {journal} {JHEP}\ }\textbf {\bibinfo {volume} {02}},\ \bibinfo {pages} {066} (\bibinfo {year} {2024}{\natexlab{b}})},\ \Eprint {http://arxiv.org/abs/2308.16125} {arXiv:2308.16125 [nucl-ex]} \BibitemShut {NoStop}%
\bibitem [{\citenamefont {Navas}\ \emph {et~al.}(2024)\citenamefont {Navas} \emph {et~al.}}]{ParticleDataGroup:2024cfk}%
  \BibitemOpen
  \bibfield  {author} {\bibinfo {author} {\bibfnamefont {S.}~\bibnamefont {Navas}} \emph {et~al.} (\bibinfo {collaboration} {Particle Data Group}),\ }\href {\doibase 10.1103/PhysRevD.110.030001} {\bibfield  {journal} {\bibinfo  {journal} {Phys. Rev. D}\ }\textbf {\bibinfo {volume} {110}},\ \bibinfo {pages} {030001} (\bibinfo {year} {2024})}\BibitemShut {NoStop}%
\bibitem [{\citenamefont {Acharya}\ \emph {et~al.}(2022{\natexlab{b}})\citenamefont {Acharya} \emph {et~al.}}]{ALICE:2022wwr}%
  \BibitemOpen
  \bibfield  {author} {\bibinfo {author} {\bibfnamefont {S.}~\bibnamefont {Acharya}} \emph {et~al.} (\bibinfo {collaboration} {ALICE}),\ }\href@noop {} {\  (\bibinfo {year} {2022}{\natexlab{b}})},\ \Eprint {http://arxiv.org/abs/2211.02491} {arXiv:2211.02491 [physics.ins-det]} \BibitemShut {NoStop}%
\end{thebibliography}%

\end{document}